\documentclass[twocolumn,english]{article}
\usepackage{tgschola}
\usepackage[LGR,T1]{fontenc}
\usepackage[latin9]{inputenc}
\usepackage{geometry}
\usepackage{babel}
\usepackage{array}
\usepackage{units}
\usepackage{textcomp}
\usepackage{multirow}
\usepackage{amsmath}
\usepackage{stmaryrd}
\usepackage{graphicx}
\usepackage[authoryear]{natbib}
\usepackage[unicode=true,pdfusetitle,
 bookmarks=true,bookmarksnumbered=false,bookmarksopen=false,
 breaklinks=false,pdfborder={0 0 0},pdfborderstyle={},backref=false,colorlinks=false]
 {hyperref}

\makeatletter

\newcommand*\LyXThinSpace{\,\hspace{0pt}}
\DeclareRobustCommand{\greektext}{%
  \fontencoding{LGR}\selectfont\def\encodingdefault{LGR}}
\DeclareRobustCommand{\textgreek}[1]{\leavevmode{\greektext #1}}
\ProvideTextCommand{\~}{LGR}[1]{\char126#1}

\providecommand{\tabularnewline}{\\}
\newcommand{\lyxaddress}[1]{
	\par {\raggedright #1
	\vspace{1.4em}
	\noindent\par}
}

\makeatother

\begin{document}
\onecolumn
\title{The planetary theory of solar activity variability: a review}
\author{Nicola Scafetta$^{1}$ and Antonio Bianchini$^{2}$}
\maketitle

\lyxaddress{$^{1}$Department of Earth Sciences, Environment and Georesources,
University of Naples Federico II, Complesso Universitario di Monte
S. Angelo, via Cintia 21, 80126 Naples, Italy.}

\lyxaddress{$^{2}$INAF, Astronomical Observatory of Padua, Padua, Italy.}

\lyxaddress{Email: nicola.scafetta@unina.it; antonio.bianchini@unipd.it}
\begin{abstract}
Commenting the 11-year sunspot cycle, \citet[MNRAS 19, 85-86]{Wolf}
conjectured that ``\emph{the variations of spot-frequency depend
on the influences of Venus, Earth, Jupiter, and Saturn}''. The high
synchronization of our planetary system is already nicely revealed
by the fact that the ratios of the planetary orbital radii are closely
related to each other through a scaling-mirror symmetry equation (Bank
and Scafetta, Front. Astron. Space Sci. 8, 758184, 2022). Reviewing
the many planetary harmonics and the orbital invariant inequalities
that characterize the planetary motions of the solar system from the
monthly to the millennial time scales, we show that they are not randomly
distributed but clearly tend to cluster around some specific values
that also match those of the main solar activity cycles. In some cases,
planetary models have even been able to predict the time-phase of
the solar oscillations including the Schwabe 11-year sunspot cycle.
We also stress that solar models based on the hypothesis that solar
activity is regulated by its internal dynamics alone have never been
able to reproduce the variety of the observed cycles. Although planetary
tidal forces are weak, we review a number of mechanisms that could
explain how the solar structure and the solar dynamo could get tuned
to the planetary motions. In particular, we discuss how the effects
of the weak tidal forces could be significantly amplified in the solar
core by an induced increase in the H-burning. Mechanisms modulating
the electromagnetic and gravitational large-scale structure of the
planetary system are also discussed.\\


\textbf{Cite as:} Scafetta, N., Bianchini, A. (2022). The planetary
theory of solar activity variability: a review. \emph{Frontiers in
Astronomy and Space Sciences, Volume 9, Article 937930.} \href{https://doi.org/10.3389/fspas.2022.937930}{https://doi.org/10.3389/fspas.2022.937930}
\end{abstract}
\twocolumn

\section{Introduction}

\begin{figure*}[!t]
\centering{}\includegraphics[width=1\textwidth]{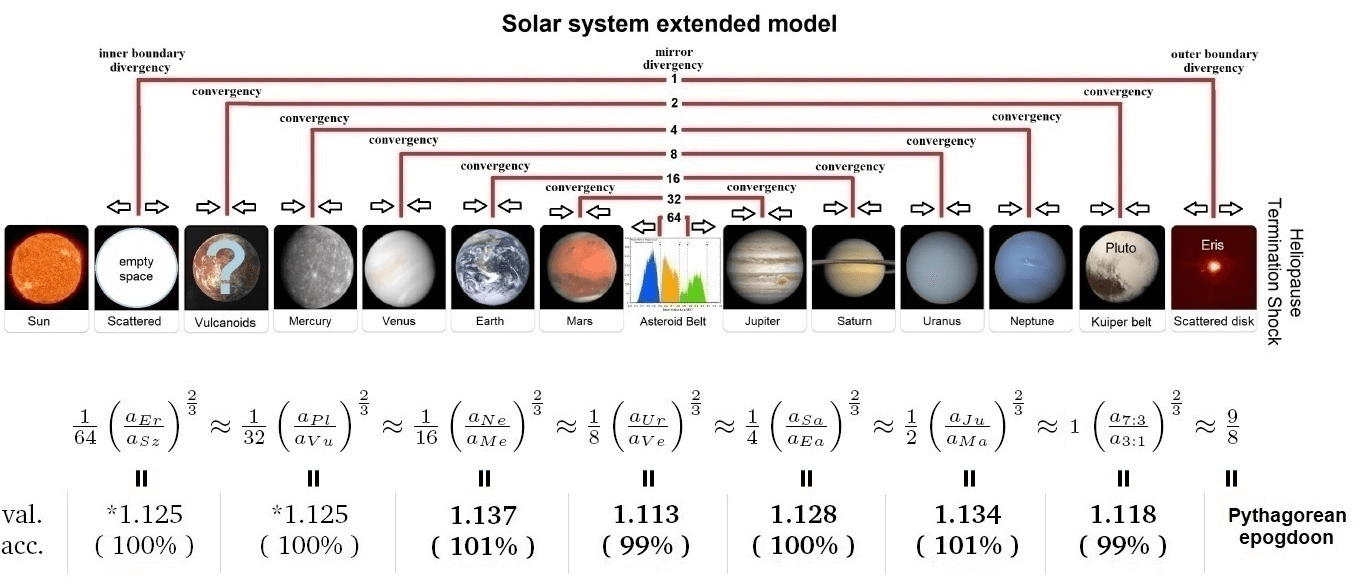}\caption{The following scaling-mirror symmetry relation among the semi-major
axes of the orbits of the planets and asteroid belts of the solar
system. \citep{Bank}.}
\label{fig0}
\end{figure*}

Since antiquity, the movements of the planets of the solar system
have attracted the attention of astronomers and philosophers such
as Pythagoras and Kepler because the orbital periods appeared to be
related to each other by simple harmonic proportions, resonances,
and/or commensurabilities \citep{Haar,Stephenson1974}. Such a philosophical
concept is known as the ``\emph{Music of the Spheres}'' or the ``\emph{Harmony
of the Worlds}'' \citep{Godwin,Scafetta2014a}. This property is
rather common for many orbital systems \citep{Agol,Aschwanden,Moons,Scafetta2014a}.
\citet{Bank} improved the Geddes and King-Hele equations describing
the mirror symmetries among the orbital radii of the planets \citep{Geddes}
and discovered their ratios obey the following scaling-mirror symmetry
relation

\begin{align}
\frac{1}{64}\left(\frac{a_{Er}}{a_{Sz}}\right)^{\frac{2}{3}} & \approx\frac{1}{32}\left(\frac{a_{Pl}}{a_{Vu}}\right)^{\frac{2}{3}}\approx\frac{1}{16}\left(\frac{a_{Ne}}{a_{Me}}\right)^{\frac{2}{3}}\approx\frac{1}{8}\left(\frac{a_{Ur}}{a_{Ve}}\right)^{\frac{2}{3}}\nonumber \\
\approx\frac{1}{4}\left(\frac{a_{Sa}}{a_{Ea}}\right)^{\frac{2}{3}} & \approx\frac{1}{2}\left(\frac{a_{Ju}}{a_{Ma}}\right)^{\frac{2}{3}}\approx1\left(\frac{a_{7:3}}{a_{3:1}}\right)^{\frac{2}{3}}\approx\frac{9}{8}\label{eq:1.0}
\end{align}
where $a_{planet}$ are the semi-major axes of the orbits of the relative
planets: Eris (Er), Pluto (Pl), Neptune (Ne), Uranus (Ur), Saturn
(Sa), Jupiter (Ju), Mars (Ma), Earth (Ea), Venus (Ve), Mercury (Me),
Vulcanoid asteroid belt (Vu), and the scattered zone surrounding the
Sun (Sz). See Figure \ref{fig0}. The ratio 9/8 is, musically speaking,
a whole tone known as the Pythagorean \emph{epogdoon}. The deviations
of Eq. \ref{eq:1.0} from the actual orbital planetary ratios are
within 1\%.

Another intriguing aspect regarding the synchronization of the solar
system is the fact that many planetary harmonics are found spectrally
coherent with the solar activity cycles \citep[e.g.: ][and many others]{Scafetta2012a,Scafetta2020}.
The precise physical origin of solar cycles is still poorly known
and dynamo models are debated, but recent literature has strengthened
the hypothesis of a correlation with planetary harmonics. Actually,
a few years after the discovery of the 11-year sunspot cycle, \citet{Wolf}
himself conjectured that ``\emph{the variations of spot-frequency
depend on the influences of Venus, Earth, Jupiter, and Saturn}''.
\citet{Dicke(1978)}  noted that the sunspot cycle shows no statistical
indication of being randomly generated but rather of being synchronized
by a chronometer hidden deep in the Sun. Solar activity is characterized
by several cycles like the Schwabe 11-year sunspot cycle \citep{Schwabe},
the Hale solar magnetic 22-year cycle \citep{Hale}, the Gleissberg
cycle ($\sim$85 years), the Jose cycle ($\sim$178 years), the Suess-de
Vries cycle ($\sim$208 years), the Eddy cycle ($\sim$1000 years),
and the Bray-Hallstatt cycle ($\sim$2300 years) \citep{Abreu,McCracken2001,McCracken2013,Scafetta2016}.
Shorter cycles are easily detected in total solar irradiance (TSI)
and sunspot records, while the longer ones are detected in long-term
geophysical records like the cosmogenic radionuclide ones (\textsuperscript{14}C
and \textsuperscript{10}Be) and  in climate records \citep{Neff,Steinhilber}.
Planetary cycles have also been found in aurora records \citep{Scafetta2012c,ScafettaWillson2013a}.

Due to the evident high synchronization of planetary motions, it is
worthwhile investigating the possibility that orbital frequencies
could tune solar variability as well. However, although Jupiter appears
to play the main role in organizing the solar system \citep{Bank},
its orbital period ($\sim$11.86 years) is too long to fit the Schwabe
11-year solar cycle. Thus, any possible planetary mechanism able to
create this solar modulation must involve a combination of more planets.
We will see that the only frequencies that could be involved in the
process are the orbital periods, the synodical periods, and their
beats and harmonics.

In the following sections, we review the planetary theory of solar
variability and show how it is today supported by many empirical and
theoretical evidences at multiple timescales. We show that appropriate
planetary harmonic models correlate with the 11-year solar cycle,
the secular and millennial cycles, as well as with several other major
oscillations observed in solar activity, and even with the occasional
occurrences of solar flares. The physics behind these results is not
yet fully understood, but a number of working hypotheses will be herein
briefly discussed.

\section{The solar dynamo and its open issues}

The hypothesis we wish to investigate is whether the solar activity
could be synchronized by harmonic planetary forcings. In principle,
this could be possible because the solar structure itself is an oscillator.
The solar cyclical magnetic activity can be explained as the result
of a dynamo operating in the convective envelope or at the interface
with the inner radiative region, where the rotational energy is converted
into magnetic energy. Under certain conditions, in particular if the
internal noise is sufficiently weak relative to the external forcing,
an oscillating system could synchronize with a weak external periodic
force, as first noted by Huygens in the 17\textsuperscript{th} century
\citep{Pikovsky}.

A comprehensive review of solar dynamo models is provided by \citet{Charbonneau(2020)}.
In the most common $\alpha$-$\Omega$ models, the magnetic field
is generated by the combined effect of differential rotation and cyclonic
convection. The mechanism starts with an initially poloidal magnetic
field that is azimuthally stretched by the differential rotation of
the convective envelope, especially at the bottom of the convective
region (tachocline) where the angular velocity gradient is most steep.
The continuous winding of the poloidal field lines ($\Omega$ mechanism)
produces a magnetic toroidal field that accumulates in the boundary
overshooting region. When the toroidal magnetic field and its magnetic
pressure get strong enough, the toroidal flux ropes become buoyantly
unstable and start rising through the convective envelope where they
undergo helical twisting by the Coriolis forces ($\alpha$ mechanism)
\citep{Parker1955}. When the twisted field lines emerge at the photosphere,
they appear as bipolar magnetic regions (BMRs), that roughly coincide
with the large sunspot pairs, also characterized by a dipole moment
that is systematically tilted with respect to the E--W direction
of the toroidal field. The turbulent decay of BMRs finally releases
a N-S oriented fraction of the dipole moment that allows the formation
of a global dipole field, characterized by a polarity reversal as
required by the observations (Babcock--Leighton mechanism).

However, magneto-hydrodynamic simulations suggest that purely interface
dynamos cannot be easily calibrated to solar observations, while flux-transport
dynamos (based on the meridional circulation) are able to better simulate
the 11-year solar cycle when the model parameters are calibrated to
minimize the difference between observed and simulated time--latitude
BMR patterns \citep{Charbonneau(2020),Dikpati}. \citet{Cole} showed
that by changing the parameters of the MHD $\alpha$-$\Omega$ dynamo
models it is possible to obtain transitions from periodic to chaotic
states via multiple periodic solutions. \citet{Macario-Rojas} obtained
a reference Schwabe cycle of 10.87 years, which was also empirically
found by \citet{Scafetta2012a} by analyzing the sunspot record. This
oscillation will be discussed later in the Jupiter-Saturn model of
Sections 4.2 and 6.

Full MHD dynamo models are not yet available and several crucial questions
are still open such as the stochastic and nonlinear nature of the
dynamo, the formation of flux ropes and sunspots, the regeneration
of the poloidal field, the modulation of the amplitude and period
of the solar cycles, how less massive fully convective stars with
no tachocline may still show the same relationship between the rotation
and magnetic activity, the role of meridional circulation, the origin
of Maunder-type Grand Minima, the presence of very low-frequency Rieger-type
periodicities probably connected with the presence of magneto-Rossby
waves in the solar dynamo layer below the convection zone, and other
issues \citep{Zaqarashvili2010,Zaqarashvili,Gurgenashvili}.

\section{The solar wobbling and its harmonic organization}

The complex dynamics of the planetary system can be described by a
general harmonic model. Any general function of the orbits of the
planets -- such as their barycentric distance, speed, angular momentum,
etc. -- must share a common set of frequencies with those of the
solar motion \citep[e.g.:][]{Jose,Bucha,Cionco2018,Scafetta2010}.
Instead, the amplitudes and phases associated with each constituent
harmonic depend on the specific chosen function.

\begin{figure*}[!t]
\centering{}\includegraphics[width=1\textwidth]{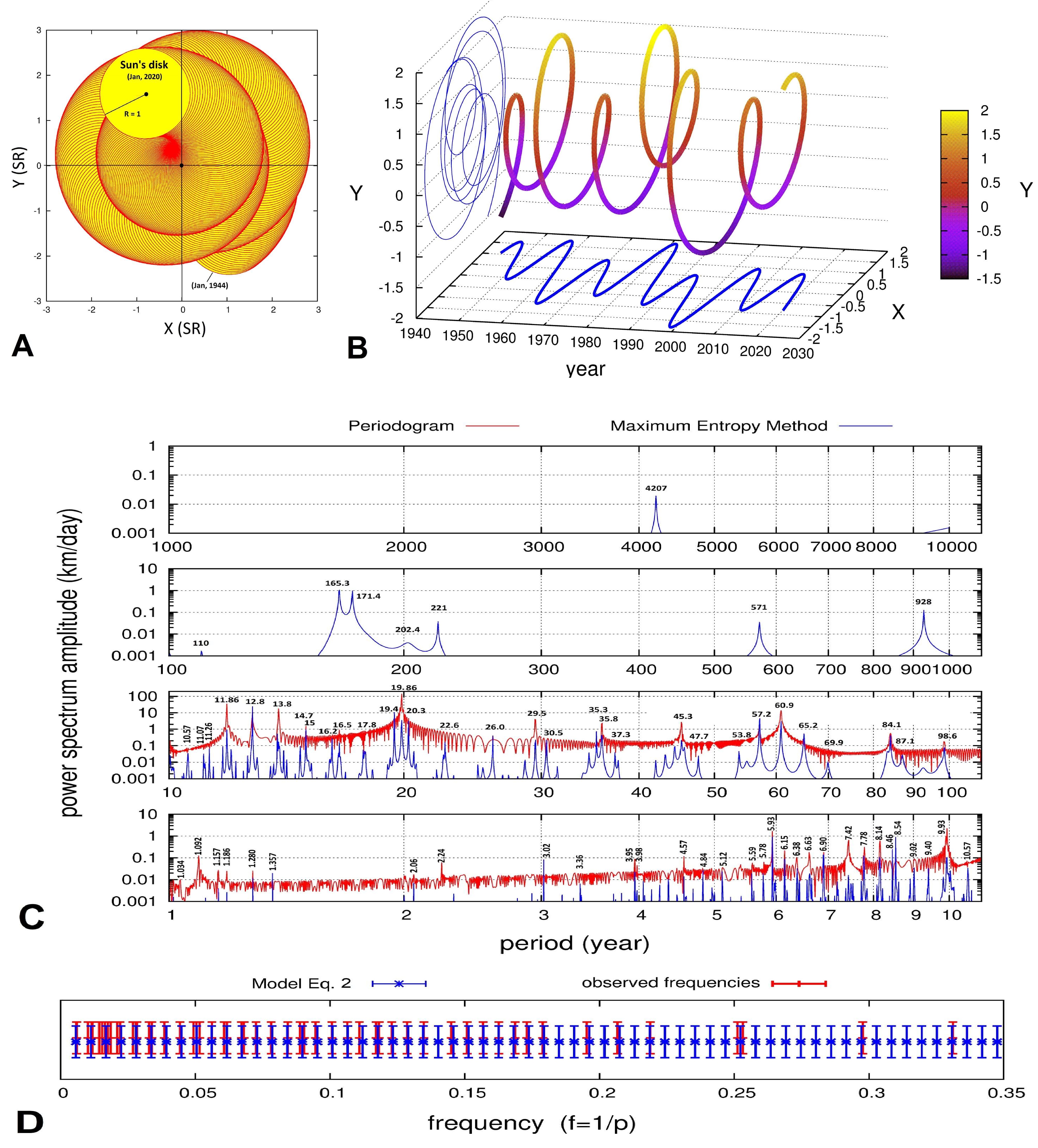}\caption{{[}A{]} The motion the wobbling Sun from 1944 to 2020 {[}B{]} The
distance and the speed of the Sun relative to the barycenter of the
solar system from 1800 to 2020. {[}C{]} Periodogram (red) and the
maximum entropy method spectrum (blue) of the speed of the Sun from
BC 8002-Dec-12, to AD 9001-Apr-24. {[}D{]} Comparison between the
frequencies observed in {[}C{]} in the range 3 to 200 years (red)
and the frequencies predicted by the harmonic model of Eq. \ref{eq:1.1}
(blue). \citep[cf.][]{Scafetta2014a}.}
\label{fig1}
\end{figure*}

Figure \ref{fig1} (A and B) shows the positions and the velocities
of the wobbling Sun with respect to the barycenter of the planetary
system from BC 8002, to AD 9001 (100-day steps) calculated using the
JPL\textquoteright s HORIZONS Ephemeris system \citep{Scafetta2010,Scafetta2014a}.

We can analyze the main orbital frequencies of the planetary system
by performing, for example, the harmonic analysis of the solar velocity
alone. Its periodograms were obtained with the Fourier analysis (red)
and the maximum entropy method (blue) \citep{Press} and are shown
in Figure \ref{fig1}C.

Several spectral peaks can be recognized: the $\sim$1.092 year period
of the Earth-Jupiter conjunctions; the $\sim$9.93 and $\sim$19.86
year periods of the Jupiter-Saturn spring (half synodic) and synodic
cycles, respectively; the $\sim$11.86, $\sim$29.5, $\sim$84 and
$\sim$165 years of the orbital periods of Jupiter, Saturn, Uranus
and Neptune, respectively; the $\sim$60 year cycle of the Trigon
of Great Conjunctions between Jupiter and Saturn; the periods corresponding
to the synodic cycles between Jupiter and Neptune ($\sim$12.8 year),
Jupiter and Uranus ($\sim$13.8 year), Saturn and Neptune ($\sim$35.8
year), Saturn and Uranus ($\sim$45.3) and Uranus and Neptune ($\sim$171.4
year), as well as their spring periods.

The synodic period is defined as

\begin{equation}
P_{12}=\frac{1}{f_{12}}=\left|\frac{1}{P_{1}}-\frac{1}{P_{2}}\right|^{-1},\label{eq:2.0}
\end{equation}
where $P_{1}$ and $P_{2}$ are the orbital periods of two planets.
Additional spectral peaks at $\sim$200-220, $\sim$571, $\sim$928
and $\sim$4200 years are also observed. The spring period is the
half of $P_{12}$. The observed orbital periods are listed in Table
\ref{tab1}.

\begin{table}[!t]
\centering{}%
\begin{tabular}{ccc}
\hline 
Planet & days & years\tabularnewline
\hline 
Mercury & 87.969 & 0.241\tabularnewline
Venus & 224.701 & 0.615\tabularnewline
Earth & 365.256 & 1\tabularnewline
Mars & 686.980 & 1.881\tabularnewline
Jupiter & 4332.589 & 11.862\tabularnewline
Saturn & 10759.22 & 29.457\tabularnewline
Uranus & 30685.4 & 84.011\tabularnewline
Neptune & 60189.0 & 164.79\tabularnewline
\hline 
\end{tabular}\caption{Sidereal orbital periods of the planets of the solar system. From
the Planetary Fact Sheet - Metric \protect\href{https://nssdc.gsfc.nasa.gov/planetary/factsheet/}{https://nssdc.gsfc.nasa.gov/planetary/factsheet/}.}
\label{tab1}
\end{table}

Some of the prominent frequencies in the power spectra appear clustered
around well-known solar cycles such as in the ranges 42-48 years,
54-70 years, 82-100 years (Gleissberg cycle), 155-185 (Jose cycle),
and 190-240 years (Suess-de Vries cycle) \citep[e.g.:][]{Ogurtsov,ScafettaWillson2013a}.
The sub-annual planetary harmonics and their spectral coherence with
satellite total solar irradiance records will be discussed in Section
5.

The important result is that the several spectral peaks observed in
the solar motion are not randomly distributed but are approximately
reproduced using the following simple empirical harmonic formula

\begin{equation}
p_{i}=\frac{178.38}{i}\quad yr,\qquad i=1,2,3,\ldots,\label{eq:1.1}
\end{equation}
 where 178 years corresponds to the period that \citet{Jose} found
both in the solar orbital motion and in the sunspot records \citep[cf.:][]{Jakubcova,Charvatova2013}.
A comparison between the observed frequencies and those predicted
by the harmonic model of Eq. \ref{eq:1.1} is shown in Figure \ref{fig1}D,
where a strong coincidence is observed. Eq. \ref{eq:1.1} suggests
that the solar planetary system is highly self-organized and synchronized.

\section{The Schwabe 11-year solar cycle}

\citet{Wolf} himself proposed that the $\sim$11-year sunspot cycle
could be produced by the combined orbital motions of Venus, Earth,
Jupiter and Saturn. In the following, we discuss two possible and
complementary solar-planetary models made with the orbital periods
of these four planets.

\subsection{The Venus-Earth-Jupiter model}

The first model relates the 11-year solar cycle with the relative
orbital configurations of Venus, Earth and Jupiter, which was first
proposed by \citet{Bendandi} as recently reminded by \citet{Battistini}.
Later, \citet{Bollinger}, \citet{Hung} and others \citep[e.g.:][]{Scafetta2012c,Tattersall,Wilson,Stefani2016,Stefani2019,Stefani2021}
developed more evolved models.

This model is justified by the consideration that Venus, Earth and
Jupiter are the three major tidal planets \citep{Scafetta2012b}.
Their alignments repeat every:

\begin{equation}
\frac{1}{f_{VEJ}}=P_{VEJ}=\left(\frac{3}{P_{V}}-\frac{5}{P_{E}}+\frac{2}{P_{J}}\right)^{-1}=22.14\:yr\label{eq:2.1}
\end{equation}
where $P_{V}=224.701$ days, $P_{E}=365.256$ days and $P_{J}=4332.589$
days are the sidereal orbital periods of Venus, Earth and Jupiter,
respectively.

The calculated 22.14-year period is very close to the $\sim$22-year
Hale solar magnetic cycle. Since the Earth--Venus--Sun--Jupiter
and Sun--Venus--Earth--Jupiter configurations present equivalent
tidal potentials, the tidal cycle would have a recurrence of $11.07$
years. This period is very close to the average solar cycle length
observed since 1750 \citep{Hung,Scafetta2012a,Stefani2016}.

\citet{Vos} found evidence for a stable Schwabe cycle with a dominant
11.04-year period over a 1000-year interval which is very close to
the above 11.07 periodicity, as suggested by \citet{Stefani(2020)}.
However, the Jupiter-Saturn model also reproduces a similar Schwabe
cycle (see Sections 4.2 and 6).

Eq. \ref{eq:2.1} is an example of ``orbital invariant inequality''
\citep{Scafettaetal2016,Scafetta2020}. Section 7 explains their mathematical
property of being simultaneously and coherently seen by any region
of a differentially rotating system like the Sun. This property should
favor the synchronization of the internal solar dynamics with external
forces varying with those specific frequencies.

Eq. \ref{eq:2.1} can be rewritten in a vectorial formalism as

\begin{equation}
(3,-5,2)=3(1,-1,0)-2(0,1,-1).\label{eq:2.2}
\end{equation}
 Each vector can be interpreted a frequency where the order of its
components correspond to the arbitrary assumed order of the planets,
in this case: (Venus, Earth, Jupiter). Thus, $(3,-5,2)\equiv3/P_{V}-5/P_{E}+2/P_{J},$
$3(1,-1,0)\equiv3(1/P_{V}-1/P_{E})$ and $-2(0,1,-1)\equiv-2(1/P_{E}-1/P_{J})$.

We observe that $(1,-1,0)$ indicates the frequency of the synodic
cycle between Venus and Earth and $(0,1,-1)$ indicates the frequency
of the synodic cycle between Earth and Jupiter (Eq. \ref{eq:2.0}).
Thus, the vector $(3,-5,2)$ indicates the frequency of the beat created
by the third harmonic of the synodic cycle between Venus and Earth
and the second harmonic of the synodic cycle between Earth and Jupiter.

Eq. \ref{eq:2.2} also means that the Schwabe sunspot cycle can be
simulated by the function:

\begin{equation}
f(t)=\cos\left(2\pi\cdot2\cdot3\frac{t-t_{VE}}{P_{VE}}\right)+\cos\left(2\pi\cdot2\cdot2\frac{t-t_{EJ}}{P_{EJ}}\right),\label{eq:2.3}
\end{equation}
where $t_{VE}=2002.8327$ is the epoch of a Venus-Earth conjunction
whose period is $P_{VE}=1.59867$ years; and $t_{EJ}=2003.0887$ is
the epoch of an Earth-Jupiter conjunction whose period is $P_{EJ}=1.09207$
years. The 11.07-year beat is obtained by doubling the synodic frequencies
given in Eq. \ref{eq:2.2}.

\begin{figure*}[!t]
\centering{}\includegraphics[width=1\textwidth]{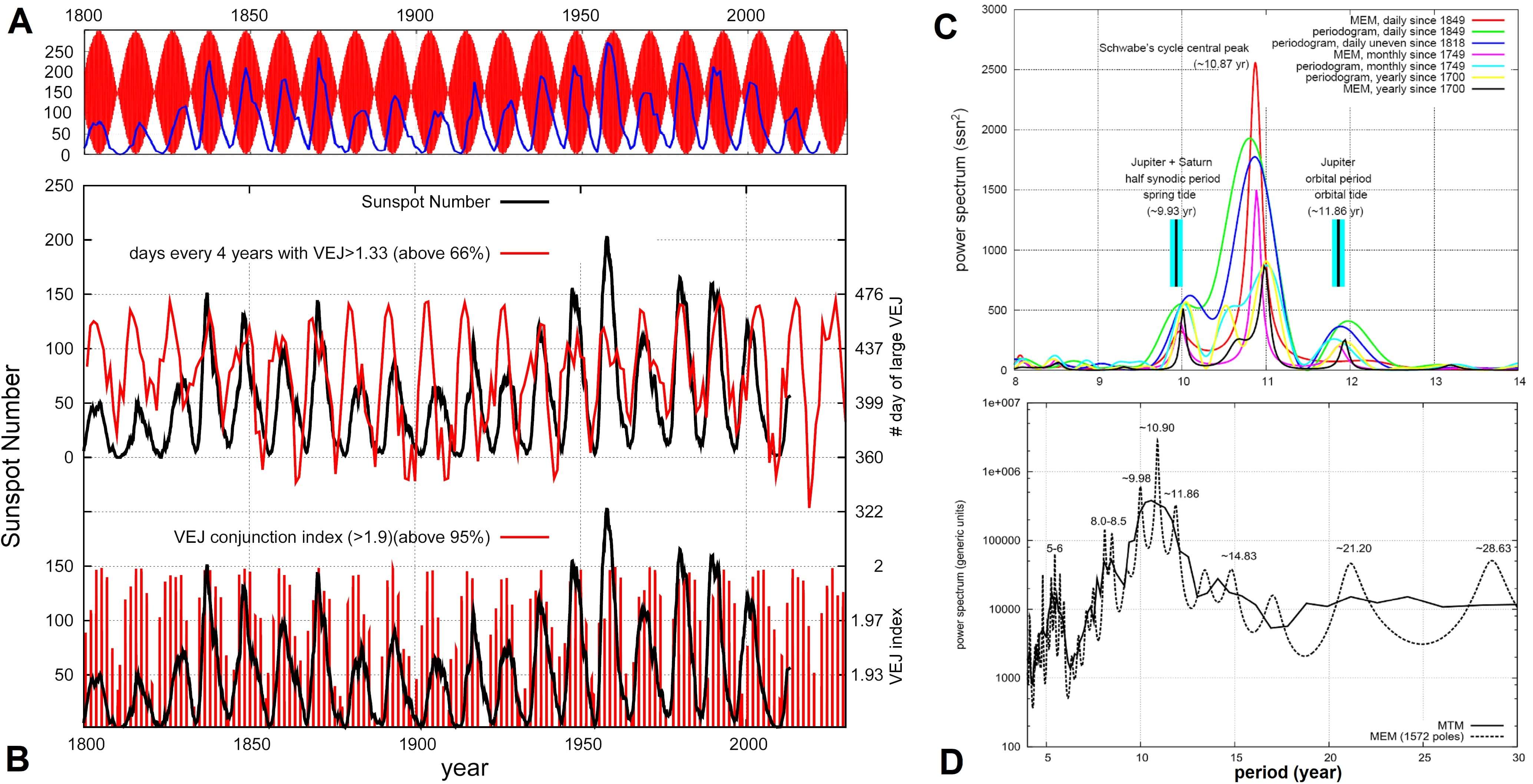}\caption{{[}A{]} The plot of Eq. \ref{eq:2.3} (red) versus the sunspot number
record (blue). {[}B, Top{]} The sunspot number record (black) versus
the alignment index $I_{VEJ}>66\%$. {[}B, Bottom{]} The sunspot number
record (black) against the number of days of most alignment ($I_{VEJ}>95\%$)
(red). {[}C and D{]} Power spectra of the Schwabe sunspot cycle using
the Maximum Entropy Method (MEM) and the periodogram (MTM) \citep{Press}.
(Data from: \protect\href{https://www.sidc.be/silso/datafiles}{https://www.sidc.be/silso/datafiles}).}
\label{fig2}
\end{figure*}

Figure \ref{fig2}A shows that the three-planet model of Eq. \ref{eq:2.3}
(red) generates a beat pattern of 11.07 years reasonably in phase
with the sunspot cycle (blue). More precisely, the maxima of the solar
cycles tend to occur when the perturbing forcing produced by the beat
is stronger, that is when the spring tides of the planets can interfere
constructively somewhere in the solar structure.

\citet{Hung} and \citet{Scafetta2012a} developed the three-planet
model by introducing a three-planetary alignment index. In the case
of two planets, the alignment index $I_{ij}$ between planet $i$
and planet $j$ is defined as:

\begin{equation}
I_{ij}=|\cos(\Theta_{ij})|,\label{eq:2.4}
\end{equation}
 where $\Theta_{ij}$ is the angle between the positions of the two
planets relative to the solar center.

Eq. \ref{eq:2.4} indicates that when the two planets are aligned
($\Theta_{ij}=0$ or $\Theta_{ij}=\pi$), the alignment index has
the largest value because these two positions imply a spring-tide
configuration. Instead, when $\Theta_{ij}=\pi/2$, the index has the
lowest value because at right angles -- corresponding to a neap-tide
configuration -- the tides of the two planets tend to cancel each
other.

In the case of the Venus-Earth-Jupiter system, there are three correspondent
alignment indexes:

\begin{eqnarray}
I_{V} & = & |\cos(\Theta_{VE})|+|\cos(\Theta_{VJ})|\label{eq:2.5}\\
I_{E} & = & |\cos(\Theta_{EV})|+|\cos(\Theta_{EJ})|\label{eq:2.6}\\
I_{J} & = & |\cos(\Theta_{JV})|+|\cos(\Theta_{JE})|.\label{eq:2.7}
\end{eqnarray}
Then, the combined alignment index $I_{VEJ}$ for the three planets
could be defined as:

\begin{equation}
I_{VEJ}=smallest~among~(I_{V},I_{E},I_{J}),\label{eq:2.8}
\end{equation}
which ranges between 0 and 2.

Figure \ref{fig2}B shows (in red) that the number of the most aligned
days of Venus, Earth and Jupiter -- estimated by Eq. \ref{eq:2.8}
-- presents an 11.07-year cycle. These cycles are well correlated,
both in phase and frequency, with the $\sim$11-year sunspot cycle.
\citet{Scafetta2012a} also showed that an 11.08-year recurrence exists
also in the amplitude and direction (latitude and longitude components)
of the solar jerk-shock vector, which is the time-derivative of the
acceleration vector. For additional details see \citet{Hung}, \citet{Scafetta2012a},
\citet{Salvador}, \citet{Wilson} and \citet{Tattersall}.

A limitation of the Venus-Earth-Jupiter model is that it cannot explain
the secular variability of the sunspot cycle which alternates prolonged
low and high activity periods such as, for example, the Maunder grand
solar minimum between 1645 and 1715, when very few sunspots were observed
\citep[cf.][]{Smythe}. However, this problem could be solved by the
Jupiter-Saturn model \citep{Scafetta2012a} discussed below and, in
general, by taking into account also the other planets \citep{Scafetta2020,Stefani2021},
as discussed in Sections 6 and 7.

The 11.07-year cycle has also been extensively studied by \citet{Stefani2016,Stefani2018,Stefani2019,Stefani2020b,Stefani2021}
where it is claimed to be the fundamental periodicity synchronizing
the solar dynamo.

\subsection{The Jupiter-Saturn model}

The second model assumes that the Schwabe sunspot cycle is generated
by the combined effects of the planetary motions of Jupiter and Saturn.
The two planets generate two main tidal oscillations associated with
the orbit of Jupiter (11.86-year period) -- which is characterized
by a relatively large eccentricity ($e=0.049$) -- and the spring
tidal oscillation generated by Jupiter and Saturn (9.93-year period)
\citep{Brown,Scafetta2012c}. In this case, the Schwabe sunspot cycle
could emerge from the synchronization of the two tides with periods
of 9.93 and 11.86 years, whose average is about 11 years.

The Jupiter-Saturn model is supported by a large number of evidences.
For example, \citet{Scafetta2012a,Scafetta2012b} showed that the
sunspot cycle length -- i.e. the time between two consecutive sunspot
minima -- is bi-modally distributed, being always characterized by
two peaks at periods smaller and larger than 11 years. This suggests
that there are two dynamical attractors at the periods of about 10
and 12 years forcing the sunspot cycle length to fall either between
10 and 11 years or between 11 and 12 years. Sunspot cycles with a
length very close to 11 years are actually absent. In addition, Figures
\ref{fig2}C and D show the periodograms of the monthly sunspot record
since 1749. The spectral analysis of this long record reveals the
presence of a broad major peak at about 10.87 years obtained by some
solar dynamo models \citep{Macario-Rojas} which is surrounded by
two minor peaks at 9.93 and 11.86 years that exactly correspond with
the two main tides of the Jupiter-Saturn system.

In Section 6 we will show that the combination of these three harmonics
produces a multidecadal, secular and millennial variability that is
rather well correlated with the long time-scale solar variability.

\section{Solar cycles shorter than the Schwabe 11-year solar cycle}

On small time scales, \citet{Bigg} found an influence of Mercury
on sunspots. Indeed, in addition to Jupiter, Mercury can also induce
relatively large tidal cycles on the Sun because its orbit has a large
eccentricity ($e=0.206$) \citep{Scafetta2012a}.

\begin{figure*}[!t]
\centering{}\includegraphics[width=1\textwidth]{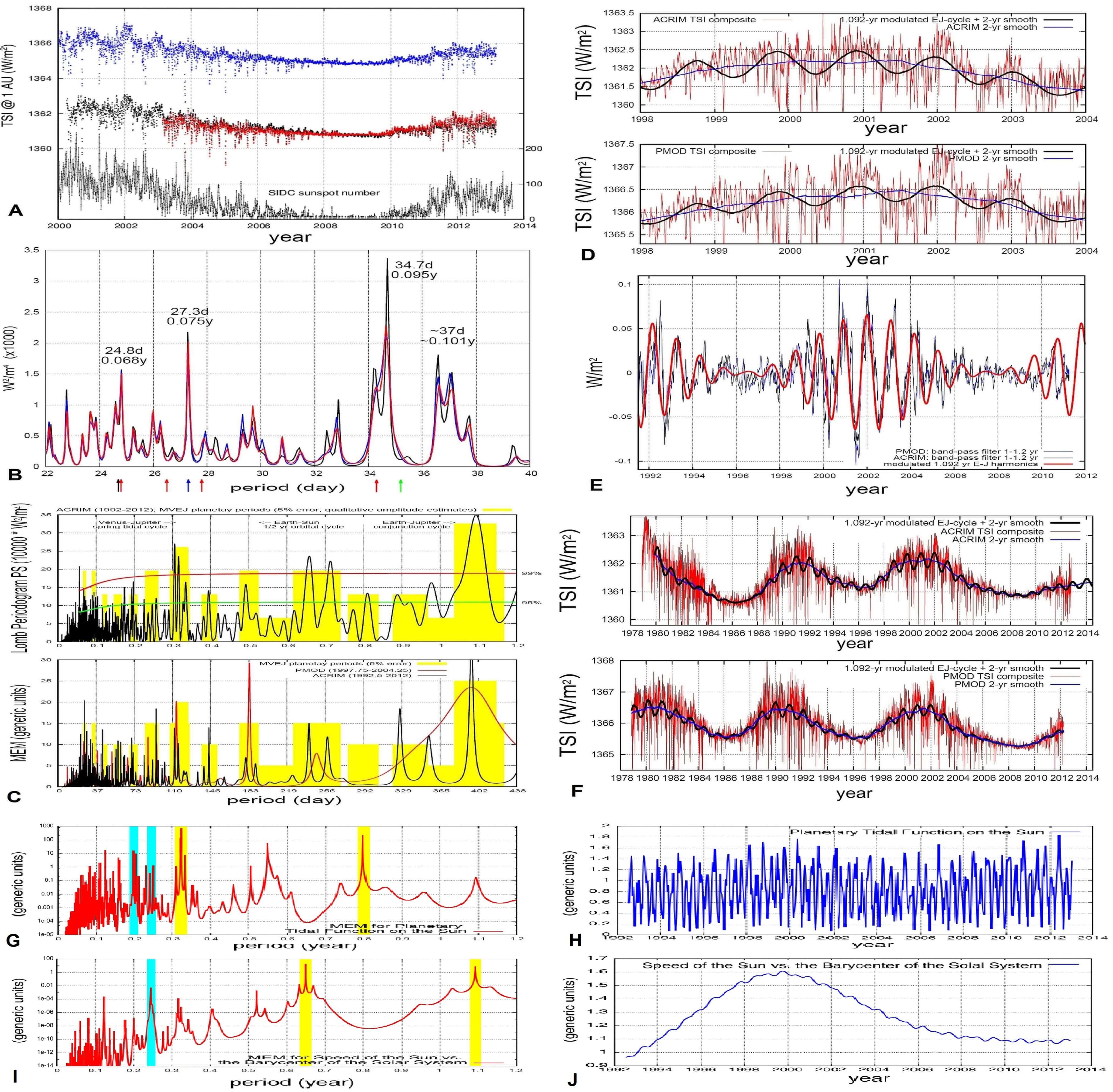}\caption{{[}A{]} Comparison of ACRIMSAT/ACRIM3 (black), SOHO/VIRGO (blue) and
SORCE/TIM (red) TSI records versus daily sunspot number (gray). {[}B{]}
Power spectrum comparison of ACRIMSAT/ACRIM3 (black), SOHO/VIRGO (blue)
and SORCE/TIM (red) TSI from 2003.15 to 2011.00. The arrows at the
bottom depicts the periods reported in Table \ref{tab3.1}. {[}C,
Top{]} Periodogram of ACRIM results in $W^{2}/m^{4}$ from 1992.5-2012.
{[}C, Bottom{]} Power spectra of ACRIM from 1992.5 to 2012.9) and
of PMOD from 1997.75 to 2004.25. The yellow bars schematically indicate
the harmonics generated by the planets as reported in Tables \ref{tab3.1},
\ref{tab3.2} and \ref{tab3.3}. {[}D{]} ACRIM and PMOD TSI composites
during solar maximum 23 (1998-2004). The black curve is from Eqs.
7 and 8. {[}E{]} High-pass filter of the PMOD (blue) and ACRIM (black)
TSI compared against a 1.092-year harmonic Jupiter function (red)..
{[}F{]} ACRIM and PMOD TSI since 1978 (red) against the models of
Eqs. \ref{eq:4.2} and \ref{eq:4.3}. {[}G, H{]} Planetary tidal function
on the Sun (blue) (see Figure \ref{fig8}C) and its power spectrum
(red). {[}I, J{]} Speed of the Sun relative to the solar system barycenter
(blue) and its power spectrum (red). \citep[cf.][]{ScafettaWillson2013b,ScafettaWillson2013c}.}
\label{fig3}
\end{figure*}

Rapid oscillations in the solar activity can be optimally studied
using the satellite total solar irradiance (TSI) records. Since 1978,
TSI data and their composites have been obtained by three main independent
science teams: ACRIMSAT/ACRIM3 \citep{Willson2003}, SOHO/VIRGO \citep{Frohlich}
and SORCE/TIM \citep{Kopp2005a,Kopp2005b}. Figure \ref{fig3} compares
the ACRIM3, VIRGO and TIM TSI from 2000 to 2014; the average irradiance
is about $1361$ $W/m^{2}$.

\subsection{The 22-40 days time-scale}

Figure \ref{fig3}B shows the power spectra in the 22-40 days range
of the three TSI records (Figure \ref{fig3}A) from 2003.15 to 2011.00
\citep{ScafettaWillson2013c}. A strong spectral peak is observed
at $\sim27.3$ days (0.075 years) \citep{Willson1999}, which corresponds
to the synodic period between the Carrington solar rotation period
of $\sim25.38$ days and the Earth's orbital period of $\sim365.25$
days. The Carrington period refers to the rotation of the Sun at $26{^\circ}$
of latitude, where most sunspots form and the solar magnetic activity
emerges \citep{Bartels}. The observed 27.3-day period differs from
the Carrington 25.38-day period because the Sun is seen from the orbiting
Earth. Thus, the 27.3-day period derives from Eq. \ref{eq:2.0} using
$T_{1}=25.38$ days and $T_{2}=365.25$ days.

Figure \ref{fig3}B reveals additional spectral peaks at $\sim24.8$
days ($\sim0.068$ years), $\sim34$-$35$ days ($\sim0.093$-$0.096$
years), and $\sim36$-$38$ days ($\sim0.099$-$0.104$ years). They
fall within the range of the solar differential rotation that varies
from 24.7-25.4 days near the equator \citep{Kotov} to about 38 days
near the poles \citep{Beck2000}.

However, the same periods appear to be also associated with the motion
of the planets. In fact, the $\sim24.8$-day cycle corresponds to
the synodic period between the sidereal orbital period of Jupiter
($\sim4332.6$ days) and the sidereal equatorial rotation period of
the Sun ($\sim24.7$ days) calculated using Eq. \ref{eq:2.0}. Additional
synodic cycles between the rotating solar equator and the orbital
motion of the terrestrial planets are calculated at $\sim26.5$ days,
relative to the Earth, $\sim27.75$ days, relative to Venus, and $\sim34.3$
days, relative to Mercury (see Table \ref{tab3.1}). We also notice
that the major TSI spectral peak at 34.7 days is very close to the
$\sim34.3$-day Mercury-Sun synodic period, although it would require
the slightly different solar rotation period of 24.89 days.

\begin{table}[!t]
\centering{}%
\begin{tabular}{ccccc}
\hline 
Cycle & Type & P (day) & P (year) & color\tabularnewline
\hline 
Sun & equ-rot & 24.7 & 0.0676 & black\tabularnewline
Sun -- Ju & equ-rot & 24.8 & 0.0679 & red\tabularnewline
Sun -- Ea & equ-rot & 26.5 & 0.0726 & red\tabularnewline
Sun -- Ea & Car-rot & 27.3 & 0.0747 & blue\tabularnewline
Sun -- Ve & equ-rot & 27.8 & 0.0761 & red\tabularnewline
Sun -- Me & equ-rot & 34.3 & 0.0940 & red\tabularnewline
2/5 Me & resonance & 35.2 & 0.0964 & green\tabularnewline
\hline 
\end{tabular}\caption{Solar equatorial (equ-) and Carrington (Car-) rotation cycles relative
to the fixed stars and to the four major tidally active planets calculated
using Eq. \ref{eq:2.0} where $P_{1}=24.7$ days is the sidereal equatorial
solar rotation and $P_{2}$ the orbital period of a planet. Last column:
color of the arrows in Figure \ref{fig3}B. \citep[cf.][]{ScafettaWillson2013c}.}
\label{tab3.1}
\end{table}

\subsection{The 0.1-1.1 year time-scale}

Tables \ref{tab3.2} and \ref{tab3.3} collect the orbital periods,
the synodic cycles and their harmonics among the terrestrial planets
(Mercury, Venus, Earth and Mars). The tables also show the synodic
cycles between the terrestrial and the Jovian planets (Jupiter, Saturn,
Uranus, and Neptune). The calculated periods are numerous and clustered.
If solar activity is modulated by planetary motions, these frequency
clusters should be observed also in the TSI records.

\begin{table*}[!t]
\centering{}%
\begin{tabular}{cccccc}
\hline 
Cycle & Type & P (day) & P (year) & min (year) & max (year)\tabularnewline
\hline 
Me & $\nicefrac{1}{2}$ orbital & $44\pm0$ & $0.1205\pm0.000$ & $0.1205$ & $0.1205$\tabularnewline
Me -- Ju & spring & $45\pm9$ & $0.123\pm0.024$ & $0.090$ & $0.156$\tabularnewline
Me -- Ea & spring & $58\pm10$ & $0.159\pm0.027$ & $0.117$ & $0.189$\tabularnewline
Me -- Ve & spring & $72\pm8$ & $0.198\pm0.021$ & $0.156$ & $0.219$\tabularnewline
Me & orbital & $88\pm0$ & $0.241\pm0.000$ & $0.241$ & $0.241$\tabularnewline
Me -- Ju & synodic & $90\pm1$ & $0.246\pm0.002$ & $0.243$ & $0.250$\tabularnewline
Ea & $\nicefrac{1}{4}$ orbital & $91\pm3$ & $0.25\pm0.000$ & $0.250$ & $0.250$\tabularnewline
Ve & $\nicefrac{1}{2}$ orbital & $112.5\pm0$ & $0.3075\pm0.000$ & $0.3075$ & $0.3075$\tabularnewline
Me -- Ea & synodic & $116\pm9$ & $0.317\pm0.024$ & $0.290$ & $0.354$\tabularnewline
Ve -- Ju & spring & $118\pm1$ & $0.324\pm0.003$ & $0.319$ & $0.328$\tabularnewline
Ea & $\nicefrac{1}{3}$ orbital & $121\pm7$ & $0.333\pm0.000$ & $0.333$ & $0.333$\tabularnewline
Me -- Ve & synodic & $145\pm12$ & $0.396\pm0.033$ & $0.342$ & $0.433$\tabularnewline
Ea & $\nicefrac{1}{2}$ orbital & $182\pm0$ & $0.500\pm0.000$ & $0.5$ & $0.5$\tabularnewline
Ea -- Ju & spring & $199\pm3$ & $0.546\pm0.010$ & $0.531$ & $0.562$\tabularnewline
Ve & orbital & $225\pm0$ & $0.615\pm0.000$ & $0.241$ & $0.241$\tabularnewline
Ve -- Ju & synodic & $237\pm1$ & $0.649\pm0.004$ & $0.642$ & $0.654$\tabularnewline
Ve -- Ea & spring & $292\pm3$ & $0.799\pm0.008$ & $0.786$ & $0.810$\tabularnewline
Ea & orbital & $365.25\pm0$ & $1.000\pm0.000$ & $1.000$ & $1.000$\tabularnewline
Ea -- Ju & synodic & $399\pm3$ & $1.092\pm0.009$ & $1.082$ & $1.104$\tabularnewline
Ea -- Ve & synodic & $584\pm6$ & $1.599\pm0.016$ & $1.572$ & $1.620$\tabularnewline
\hline 
\end{tabular}\caption{Major theoretical planetary harmonics with period $P<1.6$ years.
The synodic period is given by Eq. \ref{eq:2.0}; the spring period
is half of it. \citep[cf.][]{ScafettaWillson2013c}.}
\label{tab3.2}
\end{table*}

\begin{table*}[!t]
\centering{}%
\begin{tabular}{ccccc}
\hline 
Cycle & Type & P (year) & Type & P (year)\tabularnewline
\hline 
Me -- Ne & spring & $0.1206$ & synodic & $0.2413$\tabularnewline
Me -- Ur & spring & $0.1208$ & synodic & $0.2416$\tabularnewline
Me -- Sa & spring & $0.1215$ & synodic & $0.2429$\tabularnewline
Me -- Ma & spring & $0.1382$ & synodic & $0.2763$\tabularnewline
Ve -- Ne & spring & $0.3088$ & synodic & $0.6175$\tabularnewline
Ve -- Ur & spring & $0.3099$ & synodic & $0.6197$\tabularnewline
Ve -- Sa & spring & $0.3142$ & synodic & $0.6283$\tabularnewline
Ve -- Ma & spring & $0.4571$ & synodic & $0.9142$\tabularnewline
Ea -- Ne & spring & $0.5031$ & synodic & $1.006$\tabularnewline
Ea -- Ur & spring & $0.5060$ & synodic & $1.0121$\tabularnewline
Ea -- Sa & spring & $0.5176$ & synodic & $1.0352$\tabularnewline
Ea -- Ma & spring & $1.0676$ & synodic & $2.1352$\tabularnewline
Ma & $\nicefrac{1}{2}$ orbital & $0.9405$ & orbital & $1.8809$\tabularnewline
Ma -- Ne & spring & $0.9514$ & synodic & $1.9028$\tabularnewline
Ma -- Ur & spring & $0.9621$ & synodic & $1.9241$\tabularnewline
Ma -- Sa & spring & $1.0047$ & synodic & $2.0094$\tabularnewline
Ma -- Ju & spring & $1.1178$ & synodic & $2.2355$\tabularnewline
Ju & $\nicefrac{1}{2}$ orbital & $5.9289$ & orbital & $11.858$\tabularnewline
Ju -- Ne & spring & $6.3917$ & synodic & $12.783$\tabularnewline
Ju -- Ur & spring & $6.9067$ & synodic & $13.813$\tabularnewline
Ju -- Sa & spring & $9.9310$ & synodic & $19.862$\tabularnewline
Sa & $\nicefrac{1}{2}$ orbital & $14.712$ & orbital & $29.424$\tabularnewline
Sa -- Ne & spring & $17.935$ & synodic & $35.870$\tabularnewline
Sa -- Ur & spring & $22.680$ & synodic & $45.360$\tabularnewline
Ur & $\nicefrac{1}{2}$ orbital & $41.874$ & orbital & $83.748$\tabularnewline
Ur -- Ne & spring & $85.723$ & synodic & $171.45$\tabularnewline
Ne & $\nicefrac{1}{2}$ orbital & $81.862$ & orbital & $163.72$\tabularnewline
Me -- (Ju -- Sa) & spring & $0.122$ & synodic & $0.244$\tabularnewline
Me -- (Ea -- Ju) & spring & $0.155$ & synodic & $0.309$\tabularnewline
Ve -- (Ju -- Sa) & spring & $0.317$ & synodic & $0.635$\tabularnewline
Ea -- (Ju -- Sa) & spring & $0.527$ & synodic & $1.053$\tabularnewline
Ve -- (Ea -- Ju) & spring & $0.704$ & synodic & $1.408$\tabularnewline
\hline 
\end{tabular}\caption{Additional expected harmonics associated with planetary orbits. The
last five rows report the synodic and spring periods of Mercury, Venus
and Earth relative to the Jupiter-Saturn and Earth-Jupiter synodic
periods calculated as $P_{1(23)}=1/|1/P_{1}-|1/P_{2}-1/P_{3}||$.
\citep[cf.][]{ScafettaWillson2013b,ScafettaWillson2013c}.}
\label{tab3.3}
\end{table*}

Figure \ref{fig3}C shows two alternative power spectra of the ACRIM
and PMOD TSI records superposed to the distribution (yellow) of the
planetary frequencies reported in Tables \ref{tab3.1}, \ref{tab3.2}
and \ref{tab3.3}. The main power spectral peaks are observed at:
$\sim0.070$, $\sim0.097$, $\sim0.20$, $\sim0.25$, 0.30-0.34, $\sim0.39$,
$\sim0.55$, 0.60-0.65, 0.7-0.9, and 1.0-1.2 years.

Figure \ref{fig3}C shows that all the main spectral peaks observed
in the TSI records appear compatible with the clusters of the calculated
orbital harmonics. For example: the Mercury-Venus spring-tidal cycle
(0.20 years); the Mercury orbital cycle (0.24 years); the Venus-Jupiter
spring-tidal cycle (0.32 years); the Venus-Mercury synodic cycle (0.40
years); the Venus-Jupiter synodic cycle (0.65 years); and the Venus-Earth
spring tidal cycle (0.80 years). A 0.5-year cycle is also observed,
which could be due to the Earth crossing the solar equatorial plane
twice a year and to a latitudinal dependency of the solar luminosity.
These results are also confirmed by the power spectra of the planetary
tidal function on the Sun (see Figure \ref{fig8}C) and of the speed
of the Sun relative to the solar system barycenter (Figures \ref{fig3}G-J).

The 1.0-1.2 year band observed in the TSI records correlates well
with the 1.092-year Earth-Jupiter synodic cycle. Actually, the TSI
records present maxima in the proximity of the Earth-Jupiter conjunction
epochs \citep{ScafettaWillson2013b}.

Figure \ref{fig3}D shows the ACRIM and PMOD TSI records (red curves)
plotted against the Earth-Jupiter conjunction cycles with the period
of 1.092 years (black curve) from 1998 to 2004. TSI peaks are observed
around the times of the conjunctions. The largest peak occurs at the
beginning of 2002 when the conjunction occurred at a minimum of the
angular separation between Earth and Jupiter (0° 13' 19\textquotedbl ).

Figure \ref{fig3}E shows the PMOD (blue) and ACRIM (black) records
band-pass filtered to highlight the 1.0-1.2 year modulation. The two
curves (blue and black) are compared to the 1.092-year harmonic function
(red):

\begin{equation}
f(t)=g(t)\cos\left[2\pi~\frac{(t-2002)}{1.09208}\right],\label{eq:4.1}
\end{equation}
where the amplitude $g(t)$ was modulated according to the observed
Schwabe solar cycle. The time-phase of the oscillation is chosen at
$t_{EJ}=2002$ because one of the Earth-Jupiter conjunctions occurred
on the 1\textsuperscript{st} of January, 2002. The average Earth-Jupiter
synodic period is 1.09208 years. The TSI 1.0-1.2 year oscillation
is significantly attenuated during solar minima (1995-1997 and 2007-2009)
and increases during solar maxima. In particular, the figure shows
the maximum of solar cycle 23 and part of the maxima of cycles 22
and 24 and confirms that the TSI modulation is well correlated with
the 1.092-year Earth-Jupiter conjunction cycle.

Figure \ref{fig3}F extends the model prediction back to 1978. Here
the TSI records are empirically compared against the following equations:
\\
for ACRIM,

\begin{equation}
f(t)=S_{A}(t)+0.2(S_{A}(t)-1360.58)\cos\left[2\pi~\frac{(t-2002)}{1.09208}\right];\label{eq:4.2}
\end{equation}
for PMOD,

\begin{equation}
f(t)=S_{P}(t)+0.2(S_{P}(t)-1365.3)\cos\left[2\pi~\frac{(t-2002)}{1.09208}\right].\label{eq:4.3}
\end{equation}
 The blue curves are the 2-year moving averages, $S_{A}(t)$ and $S_{P}(t)$,
of the ACRIM and PMOD TSI composite records, respectively. The data-model
comparison confirms that the 1.092-year Earth-Jupiter conjunction
cycle is present since 1978. In fact, TSI peaks are also found in
coincidences with a number of Earth-Jupiter conjunction epochs like
those of 1979, 1981, 1984, 1990, 1991, 1992, 1993, 1994, 1995, 1998,
2011 and 2012. The 1979 and 1990 peaks are less evident in the PMOD
TSI record, likely because of the significant modifications of the
published Nimbus7/ERB TSI record in 1979 and 1989-1990 proposed by
the PMOD science team \citep{Frohlich,Scafetta2009,Scafetta2011}.

The result suggests that the side of the Sun facing Jupiter could
be slightly brighter, in particular during solar maxima. Thus, when
the Earth crosses the Sun-Jupiter line, it could receive an enhanced
amount of radiation. This coalesces with strong hotspots observed
on other stars with orbiting close giant planets \citep{Shkolnik2003,Shkolnik2005}.
Moreover, \citet{KotovH} analyzed 45 years of observations and showed
that the solar photosphere, as seen from the Earth, is pulsating with
two fast and relatively stable periods $P_{0}=9,600.606(12)$ s and
$P_{1}=9,597.924(13)$ s. Their beatings occur with a period of 397.7(2.6)
days, which coincides well with the synodic period between Earth and
Jupiter (398.9\LyXThinSpace days). A hypothesis was advanced that
the gravity field of Jupiter could be involved in the process.

\subsection{The solar cycles in the 2-9 year range}

The power spectrum in Figure \ref{fig2}D shows peaks at 5-6 and 8.0-8.5
years. The former ones appear to be harmonics of the Schwabe 11-year
solar cycle discussed in Section 3. The latter peaks are more difficult
to be identified. In any case, some planetary harmonics involving
Mercury, Venus, Earth, Jupiter and Saturn could explain them.

For example, the Mercury-Venus orbital combination repeats almost
every 11.08 years, which is similar to the 11.07-year invariant inequality
between Venus, Earth and Jupiter discussed in Section 3. In fact,
$P_{M}=0.241$ years and $P_{V}=0.615$ years, therefore their closest
geometrical recurrences occur after 23 orbits of Mercury ($23P_{M}=5.542$
years) and 9 orbits of Venus ($9P_{V}=5.535$ year). Moreover, we
have $46P_{M}=11.086$ years and $18P_{V}=11.07$ years. Thus, the
orbital configuration of Mercury and Venus repeats every 5.54 years
as well as every 11.08 years and might contribute to explain the 5-6
years spectral peak observed in Figure \ref{fig2}D. Moreover, 8 orbits
of the Earth ($8P_{E}=8$ years) and 13 orbits of Venus ($13P_{V}=7.995$
years) nearly coincide and this combination might have contributed
to produce the spectral peak at about 8 years.

There is also the possibility that the harmonics at about 5.5 and
8-9 years could emerge from the orbital combinations of Venus, Earth,
Jupiter and Saturn. In fact, we have the following orbital invariant
inequalities

\begin{equation}
\left(\frac{2}{P_{V}}-\frac{3}{P_{E}}-\frac{2}{P_{J}}+\frac{3}{P_{S}}\right)^{-1}=5.43\:yr\label{eq:4.4}
\end{equation}
and

\begin{equation}
2\left(-\frac{1}{P_{V}}+\frac{2}{P_{E}}-\frac{2}{P_{J}}+\frac{1}{P_{S}}\right)^{-1}=8.34\:yr,\label{eq:4.5}
\end{equation}
where the orbital periods of the four planets are given in Table \ref{tab1}.
Eq. \ref{eq:4.4} combines the spring cycle between Venus and Jupiter
with the third harmonic of the synodic cycle between Earth and Saturn.
Eq. \ref{eq:4.5} is the first inferior harmonic (because of the factor
2) of a combination of the synodic cycle between Venus and Saturn
and the spring cycle between Earth and Jupiter. Eqs. \ref{eq:4.4}
and \ref{eq:4.5} express orbital invariant inequalities, whose general
physical properties are discussed in Section 7.

The above results, together with those discussed in Section 4, once
again suggest that the major features of solar variability at the
decadal scale from 2 to 22 years could have been mostly determined
by the combined effect of Venus, Earth, Jupiter and Saturn, as it
was first speculated by \citet{Wolf}.

\section{The multi-decadal and millennial solar cycles predicted by the Jupiter-Saturn
model}

As discussed in Section 4.1, the Jupiter-Saturn model interprets quite
well two of the three main periods that characterize the sunspot number
record since 1749: $P_{S1}=9.93$, $P_{S2}=10.87$ and $P_{S3}=11.86$
years (Figure \ref{fig2}C) \citep{Scafetta2012a}. The two side frequencies
match the spring tidal period of Jupiter and Saturn (9.93 years),
and the tidal sidereal period of Jupiter (11.86 years). The central
peak at $P_{S2}=10.87$ years can be associated with a possible natural
dynamo frequency that is also predicted by a flux-transport dynamo
model \citep{Macario-Rojas}. However, the same periodicity could
be also interpreted as twice the invariant inequality period of Eq.
\ref{eq:4.4}, which gives 10.86 years. According to the latter interpretation,
the central frequency sunspot peak might derive from a dynamo synchronized
by a combination of the orbital motions of Venus, Earth, Jupiter and
Saturn.

The three harmonics of the Schwabe frequency band beat at $P_{S13}=60.95$
years, $P_{S12}=114.78$ years and $P_{S23}=129.95$ years. Using
the same vectorial formalism introduced in Section 3.1 to indicate
combinations of synodical cycles, a millennial cycle, $P_{S123}$,
is generated by the beat between $P_{S12}\equiv(1,-1,0)$ and $P_{S23}\equiv(0,1,-1)$
according to the equation $(1,-1,0)-(0,1,-1)=(1,-2,1)$ that corresponds
to the period

\begin{equation}
P_{S123}=\left(\frac{1}{P_{S1}}-\frac{2}{P_{S2}}+\frac{1}{P_{S3}}\right)^{-1}\approx983~yr,\label{eq:4.6}
\end{equation}
where we adopted the multi-digits accurate values $P_{S1}=9.929656$
years, $P_{S2}=10.87$ years and $P_{S3}=11.862242$ years (Table
\ref{tab1}). However, the millennial beat is very sensitive to the
choice of $P_{S2}$.

To test whether this three-frequency model actually fits solar data,
\citet{Scafetta2012a} constructed its constituent harmonic functions
by setting their relative amplitudes proportional to the power of
the spectral peaks of the sunspot periodogram. The three amplitudes,
normalized with respect to $A_{S2}$, are: $A_{S1}=0.83$, $A_{S2}=1$,
$A_{S3}=0.55$.

The time-phases of the two side harmonics are referred to: $t_{S1}=2000.475$,
which is the synodic conjunction epoch of Jupiter and Saturn (23/June/2000)
relative to the Sun, when the spring tide must be stronger; and $t_{S3}=1999.381$,
which is the perihelion date of Jupiter (20/May/1999) when its tide
is stronger. The time-phase of the central harmonic was set to $t_{S2}=2002.364$
and was estimated by fitting the sunspot number record with the three-harmonic
model keeping the other parameters fixed.

The time-phases of the beat functions are calculated using the equation
\begin{equation}
t_{12}=\frac{P_{2}t_{1}-P_{1}t_{2}}{P_{2}-P_{1}}~.\label{eq:4.7}
\end{equation}
It was found $t_{S12}=2095.311$, $t_{S13}=2067.044$ and $t_{S23}=2035.043$.
The time-phase of the beat between $P_{S12}$ and $P_{S23}$ was calculated
as $t_{S123}=2059.686$. Herein, we ignore that the phases for the
conjunction of Jupiter and Saturn vary by a few months from the average
because the orbits are elliptic, which could imply a variation up
to a few years of the time phases of the beat functions.

\begin{figure*}[!t]
\centering{}\includegraphics[width=1\textwidth,height=0.8\textheight]{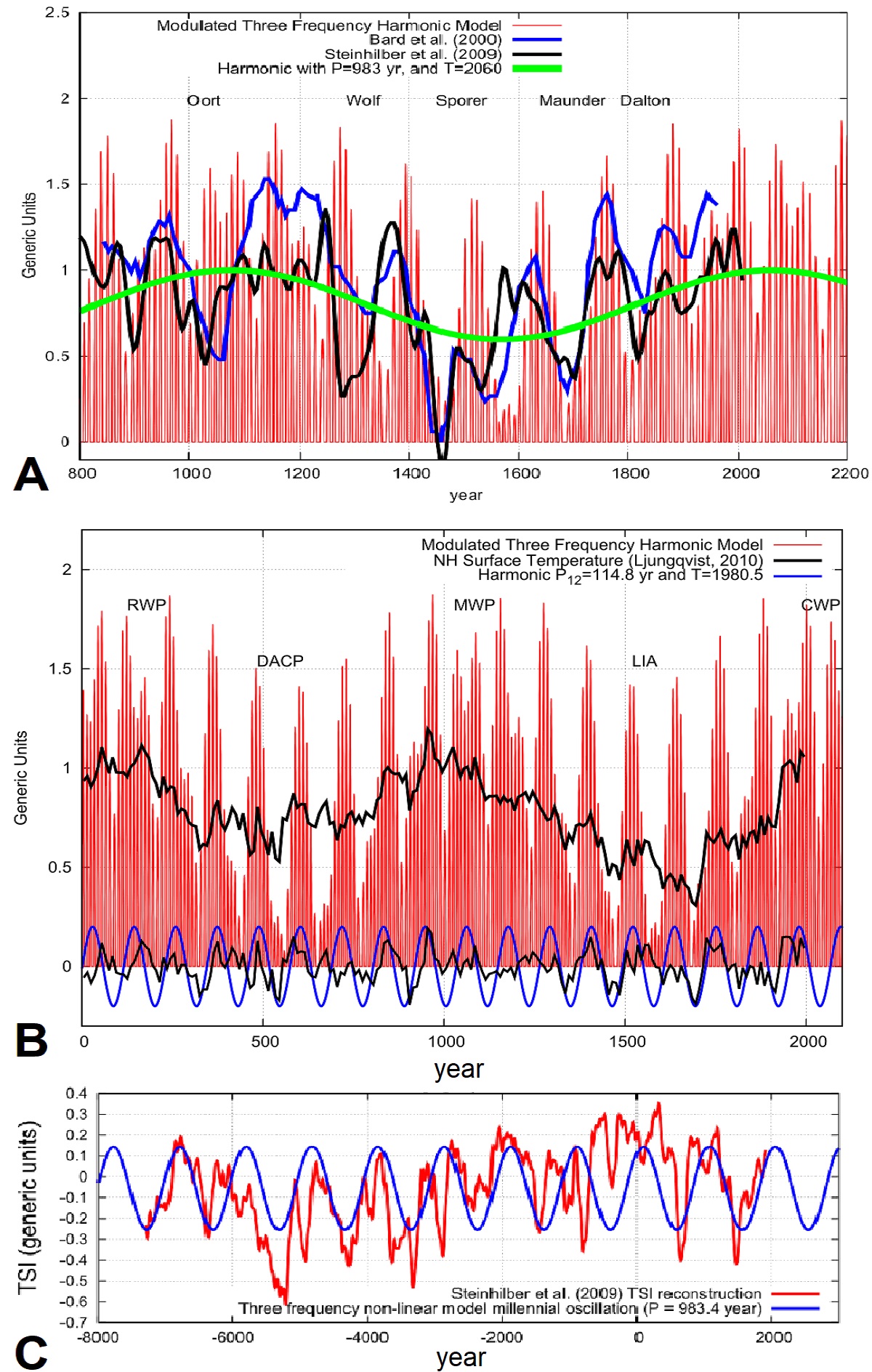}
\caption{{[}A{]} Eq. \ref{eq:5.12} (red) against two reconstructions of solar
activity based on $^{10}$Be and $^{14}$C cosmogenic isotopes \citep{Bard,Steinhilber}.
{[}B{]}. Eq. \ref{eq:5.12} (red) against a Northern Hemisphere proxy
temperature reconstruction by \citet{Ljungqvist}. {[}C{]} The millennial
oscillation predicted by the three-frequency non-linear solar model
(blue) versus the TSI proxy model by \citet{Steinhilber} (red). \citep[cf.][]{Scafetta2012a,Scafetta2014b}.}
\label{fig4}
\end{figure*}

The proposed three-frequency harmonic model is then given by the function

\begin{equation}
\sum_{i=1}^{3}h_{i}(t)=\sum_{i=1}^{3}A_{Si}~\cos\left(2\pi~\frac{t-t_{Si}}{P_{Si}}\right).\label{eq:4.8}
\end{equation}
The components and the beat functions generated by the model are given
by the equations

\begin{equation}
h_{1}(t)=0.83~\cos\left(2\pi~\frac{t-2000.475}{9.929656}\right),\label{eq:4.9}
\end{equation}

\begin{equation}
h_{2}(t)=1.0~\cos\left(2\pi~\frac{t-2002.364}{10.87}\right),\label{eq:4.10}
\end{equation}

\begin{equation}
h_{3}(t)=0.55~\cos\left(2\pi~\frac{t-1999.381}{11.862242}\right).\label{eq:4.11}
\end{equation}
Thus, the final model becomes

\begin{equation}
h_{123}(t)=h_{1}(t)+h_{2}(t)+h_{3}(t).\label{eq:4.12}
\end{equation}
To emphasize its beats we can also write

\begin{equation}
f_{123}(t)=\begin{cases}
h_{123}(t) & \text{if }h_{123}(t)\geq0\\
0 & \text{if }h_{123}(t)<0
\end{cases}\label{eq:4.13}
\end{equation}
The resulting envelope functions of the beats are

\begin{equation}
b_{12}(t)=0.60~\cos\left(2\pi~\frac{t-1980.528}{114.783}\right)\label{eq:4.14}
\end{equation}

\begin{equation}
b_{13}(t)=0.40~\cos\left(2\pi~\frac{t-2067.044}{60.9484}\right)\label{eq:4.15}
\end{equation}

\begin{equation}
b_{23}(t)=0.45~\cos\left(2\pi~\frac{t-2035.043}{129.951}\right)\label{eq:4.16}
\end{equation}

Figure \ref{fig4} shows the three-frequency solar model of Eq. \ref{eq:4.13}
(red). Figure \ref{fig4}A compares it against two reconstructions
of the solar activity based on $^{10}$Be and $^{14}$C cosmogenic
isotopes (blue and black, respectively) \citep{Bard,Steinhilber}.
The millennial beat cycle is represented by the green curve. The model
correctly hindcast all solar multi-decadal grand minima observed during
the last 1000 years, known as the Oort, Wolf, Spörer, Maunder and
Dalton grand solar minima. They approximately occurred when the three
harmonics interfered destructively. Instead, the multi-decadal grand
maxima occurred when the three harmonics interfere constructively
generating a larger perturbation on the Sun.

Figure \ref{fig4}B compares Eq. \ref{eq:4.13} against the Northern
Hemisphere proxy temperature reconstruction of \citet{Ljungqvist}
(black). We notice the good time-matching between the oscillations
of the model and the temperature record of both the millennial and
the 115-year modulations, which is better highlighted by the smoothed
filtered curves at the bottom of the figure. The Roman Warm Period
(RWP), Dark Age Cold Period (DACP), Medieval Warm Period (MWP), Little
Ice Age (LIA) and the Current Warm Period (CWP) are well hindcast
by the three-frequency Jupiter-Saturn model.

Figure \ref{fig4}C shows the millennial oscillation (blue) predicted
by Eq. \ref{eq:4.13} given by

\begin{equation}
g_{m}(t)=\cos\left(2\pi~\frac{t-2059.686}{983.401}\right).\label{eq.4.17}
\end{equation}
The curve is well correlated with the quasi millennial solar oscillation
-- known as the Eddy oscillation -- throughout the Holocene as revealed
by the $^{14}$C cosmogenic isotope record (red) and other geological
records \citep{Kerr,Scafetta2012a,Scafetta2014b,Steinhilber}.

\citet{Scafetta2012a} discussed other properties of the three-frequency
solar model. For example, five 59-63 year cycles appear in the period
1850-2150, which are also well correlated with the global surface
temperature maxima around about 1880, 1940 and 2000. The model also
predicts a grand solar minimum around the 2030s constrained between
two grand solar maxima around 2000 and 2060. The modeled solar minimum
around 1970, the maximum around 2000 and the following solar activity
decrease, which is predicted to last until the 2030s, are compatible
with the multidecadal trends of the ACRIM TSI record \citep{Willson2003},
but not with those shown by the PMOD one \citep{Frohlich} that uses
TSI modified data \citep{Scafetta2019b} and has a continuous TSI
decrease since 1980. The plots of ACRIM and PMOD TSI data are shown
in Figure \ref{fig3}F and have been extensively commented by \citet{Scafetta2019b}.
Finally, the model also reproduces a rather long Schwabe solar cycle
of about 15 years between 1680 and 1700. This long cycle was actually
observed both in the $\delta^{18}O$ isotopic concentrations found
in Japanese tree rings (a proxy for temperature changes) and in $^{14}$C
records (a proxy for solar activity) \citep{Yamaguchia}.

\citet{Scafetta2014b} also suggested that the input of the planetary
forcing could be nonlinearly processed by the internal solar dynamo
mechanisms. As a consequence, the output function might be characterized
by additional multi-decadal and secular harmonics. The main two frequency
clusters are predicted at 57, 61, 65 years and at 103, 115, 130, and
150 years. These harmonics actually appear in the power spectra of
solar activity \citep{Ogurtsov}. In particular, \citet{Cauquoin}
found the four secular periods (103, 115, 130, 150 years) in the $^{10}Be$
record of 325--336 kyr ago. These authors claimed that their analyzed
records do not show any evidence of a planetary influence but they
did not realize that their found oscillations could be derived from
the beating among the harmonics of Jupiter and Saturn with the 11-year
solar cycle, as demonstrated in \citet{Scafetta2014b}.

We notice that the multi-secular and millennial hindcasts of the solar
activity records made by the three-frequency Jupiter-Saturn model
shown in Figure \ref{fig4} are impressive because the frequencies,
phases and amplitudes of the model are theoretically deduced from
the orbits of Jupiter and Saturn and empirically obtained from the
sunspot record from 1750 to 2010. The prolonged periods of high and
low solar activity derive from the constructive and destructive interference
of the three harmonics.

\section{Orbital invariant inequality model: the Jovian planets and the long
solar and climatic cycles}

The orbital invariant inequality model was first proposed by \citet{Scafettaetal2016}
and successively developed by \citet{Scafetta2020} using only the
orbital periods of the four Jovian planets (Table \ref{tab1}). It
successfully reconstructs the main solar multi-decadal to millennial
oscillations like those observed at 55-65 years, 80-100 years (Gleissberg
cycle), 155-185 years (Jose cycle), 190-240 years (Suess-de Vries
cycle), 800-1200 years (Eddy cycle) and at 2100-2500 years (Bray-Hallstatt
cycle) \citep{Abreu,McCracken2001,McCracken2013,Scafetta2016}. The
model predictions well agree with the solar and climate long-term
oscillations discussed, for example, in \citet{Neff} and \citet{McCracken2013}.
Let us now describe the invariant inequality model in some detail.

Given two harmonics with period $P_{1}$ and $P_{2}$ and two integers
$n_{1}$ and $n_{2}$, there is a resonance if $P_{1}/P_{2}=n_{1}/n_{2}$.
In the real planetary motions, this identity is almost always not
satisfied. Consequently, it is possible to define a new frequency
$f$ and period $P$ using the following equation

\begin{equation}
f=\frac{1}{P}=\left|\frac{n_{1}}{P_{1}}-\frac{n_{2}}{P_{2}}\right|,\label{eq:5.1}
\end{equation}
which is called ``inequality''. Clearly, $f$ and $P$ represent
the beat frequency and the beat period between $n_{1}/P_{1}$ and
$n_{2}/P_{2}$. The simplest case is when $n_{1}=n_{2}=1$, which
corresponds to the synodal period between two planets defined in Eq.
\ref{eq:2.0}, which is reported below for convenience:

\begin{equation}
P_{12}=\frac{1}{f_{12}}=\left|\frac{1}{P_{1}}-\frac{1}{P_{2}}\right|^{-1}.\label{eq:5.2}
\end{equation}
Eq. \ref{eq:5.2} indicates the average time interval between two
consecutive planetary conjunctions relative to the Sun. The conjunction
periods among the four Jovian planets are reported in Table \ref{tab5.1}.

\begin{table*}[!t]
\centering{}%
\begin{tabular}{cccccc}
\hline 
 & Inv. Ineq. & Period (year) & Julian Date & Date & Long.\tabularnewline
\hline 
Jup-Sat & (1,-1,0,0) & 19.8593 & 2451718.4 & 2000.4761 & 52° 01'\tabularnewline
Jup-Ura & (1,0,-1,0) & 13.8125 & 2450535.8 & 1997.2383 & 305° 22'\tabularnewline
Jup-Nep & (1,0,0,-1) & 12.7823 & 2450442.1 & 1996.9818 & 297° 21'\tabularnewline
Sat-Ura & (0,1,-1,0) & 45.3636 & 2447322.1 & 1988.4397 & 269° 05'\tabularnewline
Sat-Nep & (0,1,0,-1) & 35.8697 & 2447725.6 & 1989.5444 & 281° 14'\tabularnewline
Ura-Nep & (0,0,1,-1) & 171.393 & 2449098.1 & 1993.3021 & 289° 22'\tabularnewline
\hline 
\end{tabular}\caption{Heliocentric synodic invariant inequalities and periods with the timing
of the planetary conjunctions closest to 2000 AD. \citep[cf.][]{Scafetta2020}.}
\label{tab5.1}
\end{table*}

Eq. \ref{eq:5.1} can be further generalized for a system of $n$
orbiting bodies with periods $P_{i}$ ($i=1,2,\ldots,n$). This defines
a generic inequality, represented by the vector $(a_{1},a_{2},\ldots,a_{n})$,
as 
\begin{equation}
f=\frac{1}{P}=\left|\sum_{i=1}^{n}\frac{a_{i}}{P_{i}}\right|,\label{eq:5.3}
\end{equation}
where $a_{i}$ are positive or negative integers.

Among all the possible orbital inequalities given by Eq. \ref{eq:5.3},
there exists a small subset of them that is characterized by the condition:

\begin{equation}
\sum_{i=1}^{n}a_{i}=0.\label{eq:5.4}
\end{equation}
This special subset of frequencies is made of the synodal planetary
periods (Eq. \ref{eq:5.2}) and all the beats among them.

It is easy to verify that the condition imposed by Eq. \ref{eq:5.4}
has a very important physical meaning: it defines a set of harmonics
that are invariant with respect to any rotating system such as the
Sun and the heliosphere. Given a reference system at the center of
the Sun and rotating with period $P_{o}$, the orbital periods, or
frequencies, seen relative to it are given by

\begin{equation}
f_{i}'=\frac{1}{P_{i}'}=\frac{1}{P_{i}}-\frac{1}{P_{o}}.\label{eq:5.5}
\end{equation}
With respect to this rotating frame of reference, the orbital inequalities
among more planets are given by:

\begin{equation}
f'=\frac{1}{P'}=\left|\sum_{i=1}^{n}\frac{a_{i}}{P_{i}'}\right|=\left|\sum_{i=1}^{n}\frac{a_{i}}{P_{i}}-\frac{\sum_{i=1}^{n}a_{i}}{P_{o}}\right|.\label{eq:5.6}
\end{equation}
If the condition of Eq. \ref{eq:5.4} is imposed, we have that $f'=f$
and $P'=P$. Therefore, this specific set of orbital inequalities
remains invariant regardless of the rotating frame of reference from
which they are observed.

For example, the conjunction of two planets relative to the Sun is
an event that is observed in the same way in all rotating systems
centered in the Sun. Since the Sun is characterized by a differential
rotation that depends on its latitude, this means that all solar regions
simultaneously feel the same planetary beats, which can strongly favor
the emergence of synchronized phenomena in the Sun. Due to this physical
property, the orbital inequalities that fulfill the condition given
by Eq. \ref{eq:5.4} were labeled as ``invariant'' inequalities.

Table \ref{tab5.2} reports the orbital invariant inequalities generated
by the large planets (Jupiter, Saturn, Uranus, and Neptune) up to
some specific order. They are listed using the vectorial formalism:

\begin{equation}
f=\frac{1}{P}=(a_{1},a_{2},a_{3},a_{4}),\label{eq:5.7}
\end{equation}
where $a_{1}$ (for Jupiter), $a_{2}$ (for Saturn), $a_{3}$ (for
Uranus) and $a_{4}$ (for Neptune) are positive or negative integers
and their sum is zero (Eq. \ref{eq:5.4}).

Two order indices, $M$ and $K$, can also be used. $M$ is the maximum
value among $|a_{i}|$ and $K$ is defined as

\begin{equation}
K=\frac{1}{2}(|a_{1}|+|a_{2}|+|a_{3}|+|a_{4}|).\label{eq:5.8}
\end{equation}
Since for the invariant inequalities the condition of Eq. \ref{eq:5.4}
must hold, $K$ indicates the number of synodal frequencies between
Jovian planet pairs producing a specific orbital invariant. For example,
$K=1$ means that the invariant inequality is made of only one synodal
frequency between two planets, $K=2$ indicates that the invariant
inequality is made of two synodal frequencies, etc.

For example, the invariant inequality cycle $(1,-3,1,1)$ has $K$
= 3 and it is the beat obtained by combining the synodal cycles of
Jupiter-Saturn, Saturn-Uranus and Saturn-Neptune because it can be
decomposed into three synodal cycles like $(1,-3,1,1)=(1,-1,0,0)-(0,1,-1,0)-(0,1,0,-1)$.
In the same way, it is possible to decompose any other orbital invariant
inequality. Hence, all the beats among the synodal cycles are invariant
inequalities and can all be obtained using the periods and time phases
listed in Table \ref{tab5.1}.

\begin{figure*}[!t]
\centering{}\includegraphics[width=1\textwidth]{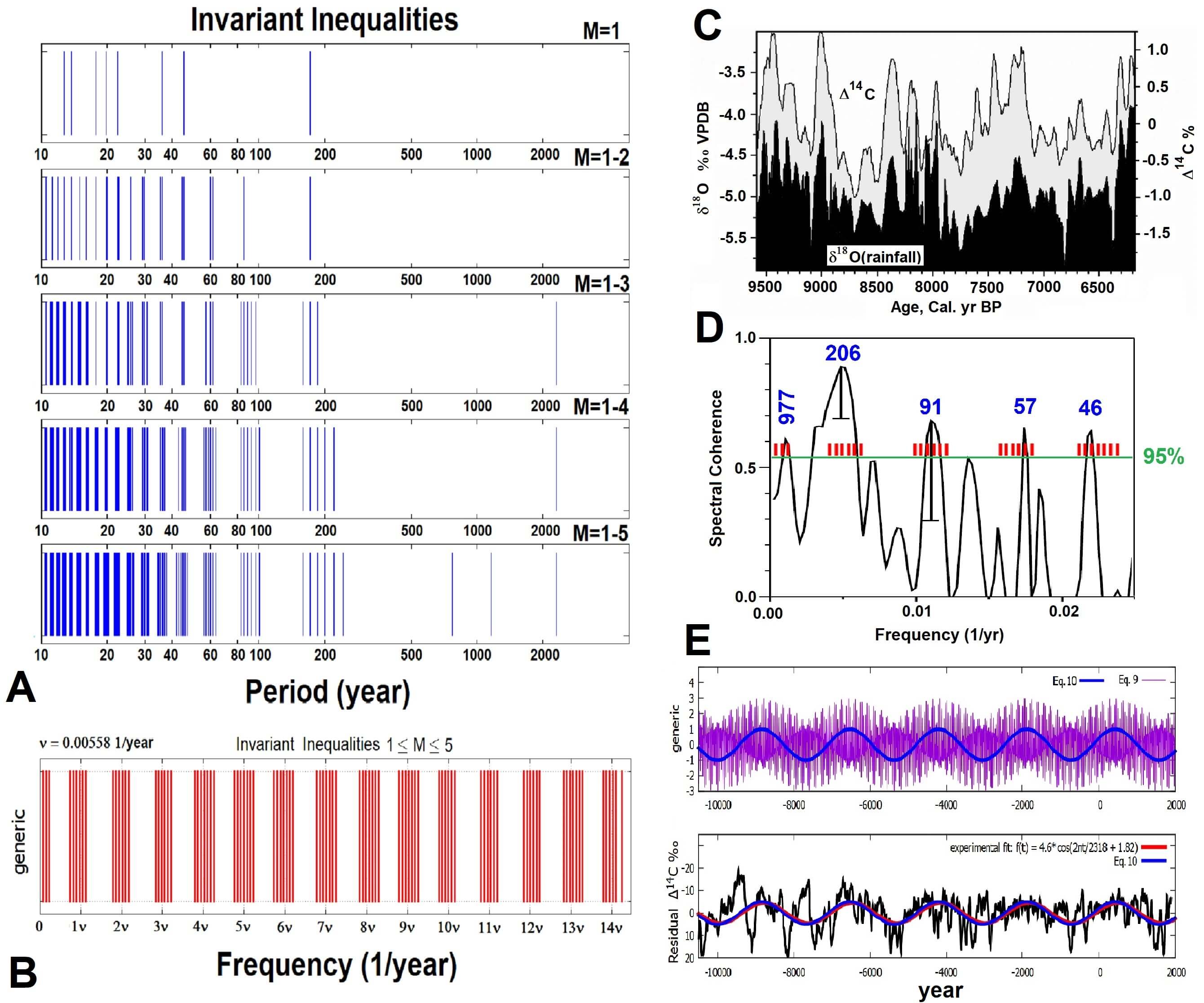}\caption{{[}A{]} The periods of the orbital invariant inequalities produced
by Jupiter, Saturn, Uranus and Neptune for $1\protect\leq M\protect\leq5$.
{[}B{]} The same harmonics highlighting their base frequency $\nu$
of the Jose cycle (179.2 years). \citep[cf.][]{Scafetta2020}. {[}C{]}
Visual correlation between the INTCAL98 atmospheric $\Delta^{14}C$
record \citep{Stuiver1998} and a speleothem calcite $\delta^{18}O$
record \citep[adapted from][]{Neff}. {[}D{]} Comparison between the
cross-spectral analysis of the two records in C against the invariant
inequalities of the solar system of Table \ref{tab5.2} (red bars).
\citep[cf.][]{Scafetta2020}. {[}E, Top{]} Eqs. \ref{eq:5.11} and
\ref{eq:5.12} that model the Hallstatt oscillation predicted by the
invariant inequality $(1,-3,1,1)$. {[}E, Bottom{]} Eq. \ref{eq:5.12}
(blue) against the the $\Delta^{14}C$ record (black) throughout the
Holocene \citep[IntCal04.14c]{Reimer} and the observed Hallstatt
oscillation deduced from a regression harmonic model (red). \citep[cf.][]{Scafettaetal2016,Scafetta2020}.}
\label{fig5}
\end{figure*}

Table \ref{tab5.2} lists all the invariant inequalities of the four
Jovian planets up to $M$ = 5. They can be collected into clusters
or groups that recall the observed solar oscillations. The same frequencies
are also shown in Figures \ref{fig5}A and B revealing a harmonic
series characterized by clusters with a base frequency of 0.00558
1/year that corresponds to the period of 179.2 years, which is known
as the Jose cycle \citeyearpar{Jose} \citep{Fairbridge,Landscheidt(1999)}.

\begin{table*}[!t]
\centering{}%
\begin{tabular}{ccccc|cccc}
\hline 
\textbf{(Jup, Sat, Ura, Nep)} &  & \textbf{(M, K)} & \textbf{T (year)} & \textbf{cluster} & \textbf{(Ven, Ear, Jup, Sat)} &  & \textbf{(M, K)} & \textbf{T (year)}\tabularnewline
\hline 
(1, -3, 5, -3) &  & (5, 6) & 42.1 &  & ( 3, -5, 5, -3 ) &  & ( 5, 8 ) & 5.10\tabularnewline
(0, 0, 4, -4) &  & (4, 4) & 42.8 &  & ( -1, 2, -3, 2 ) &  & ( 3, 4 ) & 5.28\tabularnewline
(2, -5, 1, 2) &  & (5, 5) & 43.7 &  & ( 2, -3, -2, 3 ) &  & ( 3, 5 ) & 5.43\tabularnewline
(-1, 3, 3,-5) &  & (5, 6) & 43.7 & \multirow{2}{*}{$\sim45$ year} & ( -3, 5, 2, -4 ) &  & ( 5, 7 ) & 6.40\tabularnewline
(1, -2, 0, 1) &  & (2, 2) & 44.5 &  & ( 0, 0, 3, -3 ) &  & ( 3, 3 ) & 6.62\tabularnewline
(0, 1, -1, 0) &  & (1, 1) & 45.4 &  & ( 3, -5, 4, -2 ) &  & ( 5, 7 ) & 6.86\tabularnewline
(-1, 4, -2, -1) &  & (4, 4) & 46.3 &  & ( -1, 2, -4, 3 ) &  & ( 4, 5 ) & 7.19\tabularnewline
(1, -1, -5, 5) &  & (5, 6) & 47.2 &  & ( 2, -3, -3, 4 ) &  & ( 4, 6 ) & 7.47\tabularnewline
\cline{5-5} 
(1, -3, 4, -2) &  & (4, 5) & 55.8 &  & ( -3, 5, 1, -3 ) &  & ( 5, 6 ) & 9.44\tabularnewline
(0, 0, 3, -3) &  & (3, 3) & 57.1 &  & ( 0, 0, 2, -2 ) &  & ( 2, 2 ) & 9.93\tabularnewline
(2, -5, 0, 3) &  & (5, 5) & 58.6 &  & ( 3, -5, 3, -1 ) &  & ( 5, 6 ) & 10.47\tabularnewline
(-1, 3, 2, -4) &  & (4, 5) & 58.6 & $\sim60$ year & ( -1, 2, -5, 4 ) &  & ( 5, 6 ) & 11.27\tabularnewline
(1, -2, -1, 2) &  & (2, 3) & 60.1 &  & ( 2, -3, -4, 5 ) &  & ( 5, 7 ) & 11.97\tabularnewline
(0, 1, -2, 1) &  & (2, 2) & 61.7 &  & ( -3, 5, 0, -2 ) &  & ( 5, 5 ) & 18.00\tabularnewline
(-1, 4, -3, 0) &  & (4, 4) & 63.4 &  & ( 0, 0, 1, -1 ) &  & ( 1, 1 ) & 19.86\tabularnewline
\cline{5-5} 
(1, -3, 3, -1) &  & (3, 4) & 82.6 &  & ( 3, -5, 2, 0 ) &  & ( 5, 5 ) & 22.14\tabularnewline
(0, 0, 2, -2) &  & (2, 2) & 85.7 &  & ( -3, 5, -1, -1 ) &  & ( 5, 5 ) & 192.8\tabularnewline
(2, -5, -1, 4) &  & (5, 6) & 89.0 &  &  &  &  & \tabularnewline
(-1, 3, 1, -3) &  & (3, 4) & 89.0 & Gleissberg &  &  &  & \tabularnewline
\cline{6-9} \cline{7-9} \cline{8-9} \cline{9-9} 
(1, -2, -2, 3) &  & (3, 4) & 92.5 &  & \textbf{(Mer, Ven, Ear, Jup)} &  & \textbf{(M, K)} & \textbf{T (year)}\tabularnewline
\cline{6-9} \cline{7-9} \cline{8-9} \cline{9-9} 
(0, 1, -3, 2) &  & (3, 3) & 96.4 &  & ( -2, 3, 4, -5 ) &  & (5, 7) & 6.63\tabularnewline
(-1, 4, -4, 1) &  & (4, 5) & 100.6 &  & ( 2, -4, -2, 4 ) &  & (4, 6) & 7.18\tabularnewline
\cline{5-5} 
(1, -3, 2, 0) &  & (3, 3) & 159.6 &  & ( 1, -2, -1, 2 ) &  & (2, 3) & 14.35\tabularnewline
(0, 0, 1, -1) &  & (1, 1) & 171.4 & \multirow{2}{*}{Jose} & ( 3, -5, 2, 0 ) &  & ( 5, 5 ) & 22.14\tabularnewline
(2, -5, -2, 5) &  & (5, 7) & 185.1 &  & ( 1, -5, 4, 0 ) &  & (5, 5) & 40.82\tabularnewline
(-1, 3, 0, -2) &  & (3, 3) & 185.1 &  &  &  &  & \tabularnewline
\cline{5-5} 
(1, -2, -3, 4) &  & (4, 5) & 201.1 &  &  &  &  & \tabularnewline
(0, 1, -4, 3) &  & (4, 4) & 220.2 & Suess-de Vries &  &  &  & \tabularnewline
(-1, 4, -5, 2) &  & (5, 6) & 243.4 &  &  &  &  & \tabularnewline
\cline{5-5} 
(0, -1, 5, -4) &  & (5, 5) & 772.7 & \multirow{2}{*}{Eddy} &  &  &  & \tabularnewline
(-1, 2, 4, -5) &  & (5, 6) & 1159 &  &  &  &  & \tabularnewline
\cline{5-5} 
(1, -3, 1, 1) &  & (3, 3) & 2318 & Bray-Hallstatt &  &  &  & \tabularnewline
\hline 
\end{tabular}\caption{(Left) List of invariant inequalities for periods $T\protect\geq40$
years and $M\protect\leq5$ for Jupiter, Saturn, Uranus, Neptune.
(Right) The same for Venus, Earth, Jupiter, and Saturn, and for Mercury,
Venus, Earth and Jupiter. \citep[cf.][]{Scafetta2020}.}
\label{tab5.2}
\end{table*}

The physical importance of the harmonics listed in Table \ref{tab5.2}
is shown in Figure \ref{fig5}C, which compares a solar activity reconstruction
from a \textsuperscript{14}C record, and the climatic reconstruction
from a $\delta{}^{18}O$ record covering the period from 9500 to 6000
years ago \citep{Neff}: the two records are strongly correlated.

Figure \ref{fig5}D shows that the two records present numerous common
frequencies that correspond to the cycles of Eddy (800--1200 years),
Suess-de Vries (190--240 years), Jose (155--185 years), Gleissberg
(80--100 years), the 55--65 year cluster, another cluster at 40-50
years, and some other features. Figure \ref{fig5}D also compares
the common spectral peaks of the two records against the clusters
of the invariant orbital inequalities (red bars) reported in Figure
\ref{fig5}B and listed in Table \ref{tab5.2}. The figure shows that
the orbital invariant inequality model well predicts all the principal
frequencies observed in solar and climatic data throughout the Holocene.

The efficiency of the model in hindcasting both the frequencies and
the phases of the observed solar cycles can also be more explicitly
shown. For example, the model perfectly predicts the great Bray-Hallstatt
cycle (2100-2500 years) that was studied in detail by \citet{McCracken2013}
and \citet{Scafettaetal2016}. The first step to apply the model is
to determine the constituent harmonics of the invariant inequality
$(1,-3,1,1)$. This cycle is a combination of the orbital periods
of Jupiter, Saturn, Uranus and Neptune that gives

\begin{equation}
P_{JSUN}=\frac{1}{f_{JSUN}}=\left(\frac{1}{P_{j}}-\frac{3}{P_{S}}+\frac{1}{P_{U}}+\frac{1}{P_{N}}\right)^{-1}=2317.56\:yr.\label{eq:5.9}
\end{equation}
The constituent harmonics are the synodic cycles of Jupiter-Saturn,
Saturn-Uranus and Saturn-Neptune as described by the following relation
\begin{equation}
(1,-3,1,1)=(1,-1,0,0)-(0,1,-1,0)-(0,1,0,-1).\label{eq:5.10}
\end{equation}
Thus, the invariant inequality $(1,-3,1,1)$ is the longest beat modulation
generated by the superposition of these three synodic cycles and it
can be expressed as the periodic function

\begin{equation}
f(t)=\sin\left(2\pi\frac{t-t_{JS}}{P_{JS}}\right)+\sin\left(2\pi\frac{t-t_{SU}}{P_{SU}}\right)+\sin\left(2\pi\frac{t-t_{SN}}{P_{SN}}\right)\label{eq:5.11}
\end{equation}
where $P_{ij}$ are the synodic periods and $t_{ij}$ are the correspondent
time-phases listed in Table \ref{tab5.1}.

Eq. \ref{eq:5.11} is plotted in Figure \ref{fig5}E and shows the
long beat modulation superposed to the Bray-Hallstatt period of 2318
years found in the $\Delta^{14}C$ (\textperthousand) record (black)
throughout the Holocene \citep[IntCal04.14c]{Reimer}. This beat cycle
is captured, for example, by the function:

\begin{equation}
f_{B}(t)=-\sin\left(2\pi\frac{t-t_{JS}}{P_{JS}}-2\pi\frac{t-t_{SU}}{P_{SU}}-2\pi\frac{t-t_{SN}}{P_{SN}}\right),\label{eq:5.12}
\end{equation}
whose period is 2318 years and the timing is fixed by the three conjunction
epochs and the respective synodic periods. In fact, the argument of
the above sinusoidal function is the sum of three terms that correspond
to those of Equation \ref{eq:5.10}. Equation \ref{eq:5.12} is plotted
in Figure \ref{fig5}E as the blue curve.

Three important invariant inequalities -- $(1,-3,2,0)$, $(0,0,1,-1)$
and $(-1,3,0,-2)$ -- are found within the Jose 155--185 year period
band:

\begin{equation}
P_{JSU}=\frac{1}{f_{JSU}}=\left(\frac{1}{P_{J}}-\frac{3}{P_{S}}+\frac{2}{P_{U}}\right)^{-1}=159.59\:yr,\label{eq:5:13}
\end{equation}

\begin{equation}
P_{UN}=\frac{1}{f_{UN}}=\left(\frac{1}{P_{U}}-\frac{1}{P_{N}}\right)^{-1}=171.39\:yr,\label{eq:5:14}
\end{equation}

\begin{equation}
P_{JSN}=\frac{1}{f_{JSN}}=\left(-\frac{1}{P_{J}}+\frac{3}{P_{S}}-\frac{2}{P_{N}}\right)^{-1}=185.08\:yr.\label{eq:5:15}
\end{equation}
The long beat between Eq. \ref{eq:5:14} and Eq. \ref{eq:5:13} --
that is $(0,0,1,-1)-(-1,3,0,-2)=(1,-3,1,-1)$ -- is the great Bray--Hallstatt
cycle. The fast beat between Eq. \ref{eq:5:14} and Eq. \ref{eq:5:15}
-- $(0,0,1,-1)+(-1,3,0,-2)=(-1,3,1,-3)$ -- is the Gleissberg 89-year
cycle, which also corresponds to half of the Jose period of $\sim$178
year that regulates the harmonic structure of the wobbling of the
solar motion.

Another interesting invariant inequality is $(1,-2,-1,2)=(1,0,-1,0)-2(0,1,0,-1)$,
which is a beat between the synodic period of Jupiter and Uranus (1,0,-1,0)
and the first harmonic of the synodic period of Saturn and Neptune.
The period is:

\begin{equation}
P_{JSN}=\frac{1}{f_{JSN}}=\left(\frac{1}{P_{J}}-\frac{2}{P_{S}}-\frac{1}{P_{U}}+\frac{2}{P_{N}}\right)^{-1}=60.1\:yr,\label{eq:5:16}
\end{equation}
The beat oscillation is given by the equation:

\begin{equation}
f(t)=\cos\left(2\pi\frac{t-t_{JU}}{P_{JU}}\right)+\cos\left(2\pi\cdot2\frac{t-t_{SN}}{P_{SN}}\right),\label{eq:5.17}
\end{equation}
that shows a 60.1-year beat oscillation. The pattern is found in both
solar and climate records and could be physically relevant because
the maxima of the 60-year beat occur during specific periods -- the
1880s, 1940s, and 2000s -- that were characterized by maxima in climatic
records of global surface temperatures and in several other climate
index records \citep{Agnihotri,Scafetta2013,Scafetta2014c,Wyatt}.
The 60-year oscillation was even found in the records of the historical
meteorite falls in China from AD 619 to 1943 \citep{ChangYu1981,Scafetta2019,Yu1983}.

An astronomical 60-year oscillation can be obtained in several ways.
In particular, \citet{Scafetta2010} and \citeyearpar{Scafetta2012c}
showed that it is also generated by three consecutive conjunctions
of Jupiter and Saturn since their synodic cycle is 19.86 years and
every three alignments the conjunctions occur nearly in the same constellation.
The three consecutive conjunctions are different from each other because
of the ellipticity of the orbits. The 60-year pattern has been known
since antiquity as the Trigon of the Great Conjunctions \citep{Kepler},
which also slowly rotates generating a quasi-millennial cycle known
as the Great Inequality of Jupiter and Saturn \citep{Etz,Lovett,Scafetta2012c,Wilson}.

Both the 60-year and the quasi-millennial oscillations also characterize
the evolution of the instantaneous eccentricity function of Jupiter
\citep{Scafetta2019}. The quasi millennial oscillation (the Heddy
cycle) could be related to the two orbital invariant inequalities
$(0,-1,5,-4)\equiv772.7$ years and $(-1,2,4,-5)\equiv1159$ years.
Their beat frequency being $(0,-1,5,-4)-(-1,2,4,-5)=(1,-3,1,1)\equiv2318$
years, which corresponds to the Bray--Hallstatt cycle. Their mean
frequency, instead, is $0.5(0,-1,5,-4)+0.5(-1,2,4,-5)=0.5(-1,1,9,-9)\equiv927$
years that reminds the Great Inequality cycle of Jupiter and Saturn
suggesting that this great cycle could also be generated by the beat
between the synodic period of Jupiter and Saturn, $(1,-1,0,0)$ and
the ninth harmonic of the synodic period of Uranus and Neptune, $9(0,0,1,-1)$.

The invariant inequality model can be extended to all the planets
of the solar system (see Tables \ref{tab3.2} and \ref{tab3.3} and
\ref{tab5.2}). The ordering of the frequencies according to their
physical relevance depends on the specific physical function involved
(e.g. tidal forcing, angular momentum transfer, space weather modulation,
etc.) and will be addressed in future work.

\section{The Suess-de Vries cycle (190-240 years)}

The Suess-de Vries cycle is an important secular solar oscillation
commonly found in radiocarbon records \citep{de Vries,Suess1965}.
Several recent studies have highlighted its importance \citep{Abreu,Beer,L=0000FCdecke,McCracken2013,Neff,Stefani2020b,Stefani2021,Wagner2001,Weiss(2016)}.
Its period varies between 200 and 215 years but the literature also
suggests a range between 190 and 240 years.

\citet{Stefani2021} argued that the Suess-de Vries cycle, together
with the Hale and the Gleissberg-type cycles, could emerge from the
synchronization between the 11.07-year periodic tidal forcing of the
Venus--Earth--Jupiter system and the 19.86-year periodic motion
of the Sun around the barycenter of the solar system due to Jupiter
and Saturn. This model yields a Suess-de Vries-type cycle of 193 years.

Actually, the 193-year period is the orbital invariant inequality
$(-3,5,-1,-1)=(0,0,1,-1)-(3,-5,2,0)$ where $(0,0,1,-1)$ is the synodic
cycle of Jupiter and Saturn (19.86 years) and $(3,-5,2,0)$ is the
22.14-year orbital inequality cycle of Venus, Earth and Jupiter (Eq.
\ref{eq:2.2}). We also notice that $(0,0,1,-1)+(3,-5,2,0)=(3,-5,3,-1)$
corresponds to the period of 10.47 years which is a periodicity that
has been observed in astronomical and climate records \citep{Scafetta2014b,Scafettaetal2020}.

The orbital invariant inequality model discussed in Section 7 provides
an alternative and/or complementary origin of the Suess-de Vries cycle.
In fact, the orbital invariant inequalities among Jupiter, Saturn,
Uranus and Neptune form a cluster of planetary beats with periods
between 200 and 240 years. Thus, the Suess-de Vries cycle might also
emerge as beat cycles among the orbital invariant inequalities with
periods around 60 years and those belonging to the Gleissberg frequency
band with periods around 85 years. See Table \ref{tab5.2}. In fact,
their synodic cycles would approximately be

\begin{equation}
\frac{1}{1/60-1/85}=204\:yr.
\end{equation}

It might also be speculated that the Suess-de Vries cycle originates
from a beat between the Trigon of the Great Conjuctions of Jupiter
and Saturn ($3\times19.862=59.6$ years, which is an oscillation that
mainly emerges from the synodical cycle between Jupiter and Saturn
combined with the eccentricity of the orbit of Jupiter) and the orbital
period of Uranus (84 years). In this case, we would have $1/(1/59.6-1/84)=205$
years.

The last two estimates coincide with the 205-year Suess-de Vries cycle
found in radiocarbon records by \citet{Wagner2001} and are just slightly
smaller than the 208-year cycle found in other similar recent studies
\citep{Abreu,Beer,McCracken2013,Weiss(2016)}

We notice that the natural planetary cycles that could theoretically
influence solar activity are either the orbital invariant inequality
cycles (which involve the synodic cycles among the planets assumed
to be moving on circular orbits) and the orbital cycles of the planets
themselves because the orbits are not circular but eccentric, and
their harmonics.

\section{Evidences for planetary periods in climatic records}

\begin{figure*}[!t]
\centering{}\includegraphics[width=1\textwidth]{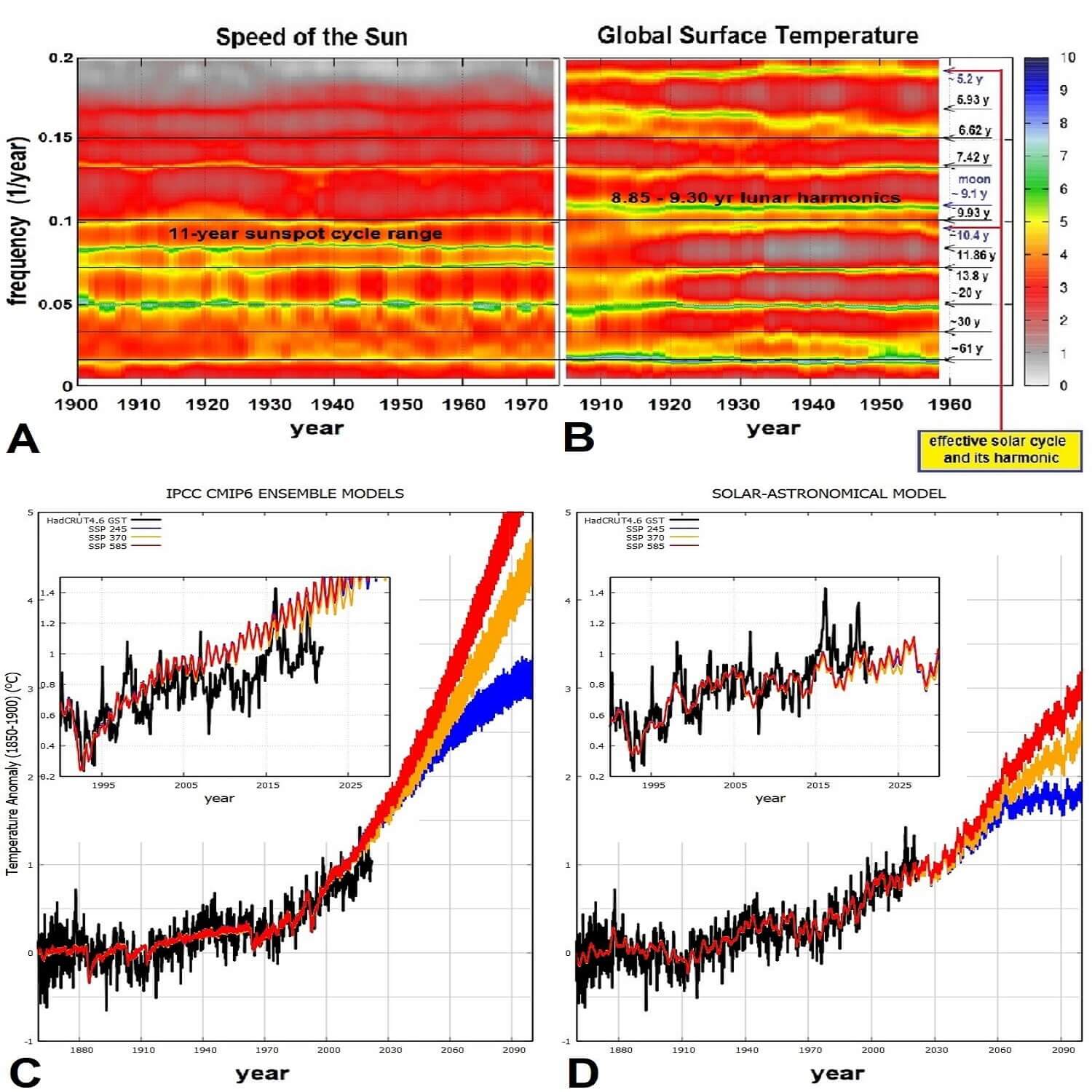}\caption{{[}A{]} Time-frequency analysis (L = 110 years) of the speed of the
Sun relative to the barycenter of the solar system. {[}B{]} Time frequency
analysis (L = 110 years) of the detrended HadCRUT3 temperature record.
\citep[cf.][]{Scafetta2014b}. {[}C{]} Ensemble CMIP6 GCM mean simulations
for different emission scenarios versus the HadCRUT global surface
temperatures. {[}D{]} The same record compared with the solar-astronomical
harmonic climate model \citet{Scafetta2013} updated in \citet{Scafetta2021}.}
\label{fig6}
\end{figure*}

A number of solar cycles match the periods found in climatic records
(see Figures \ref{fig4}, \ref{fig5} and \ref{fig6}) and often appear
closely correlated for millennia \citep[e.g.:][and many others]{Neff,Scafetta2004,Scafetta2006,Scafetta2009,Scafetta2021,Steinhilber2012}.

Evidences for a astronomical origin of the Sub-Milankovitch climate
oscillations have been discussed in several studies \citep[e.g.:][]{Scafetta2010,Scafetta2014b,Scafetta2016,Scafetta2018,Scafetta2021}.
Let us now summarizes the main findings relative to the global surface
temperature record from 1850 to 2010.

Figures \ref{fig6}A and B compare the time-frequency analyses between
the speed of the Sun relative to the center of mass of the solar system
(Figure \ref{fig1}) and the HadCRUT3 global surface records \citep{Scafetta2014b}.
It can be seen that the global surface temperature oscillations mimic
several astronomical cycles at the decadal and multidecadal scales,
as first noted in \citet{Scafetta2010} and later confirmed by advanced
spectral coherence analyses \citep{Scafetta2016,Scafetta2018}.

The main periods found in the speed of the Sun (Figure \ref{fig6}A)
are at about 5.93, 6.62, 7.42, 9.93, 11.86, 13.8, 20 and 60 years.
Most of them are related to the orbits of Jupiter and Saturn. The
main periods found in the temperature record (Figure \ref{fig6}B)
are at about 5.93, 6.62, 7.42, 9.1, 10.4, 13.8, 20 and 60 years. Most
of these periods appear to coincide with orbital invariant inequalities
(Table \ref{tab5.2}) but the 9.1 and 10.4-year cycles.

Among the climate cycles, it is also found an important period of
about 9.1 years, which is missing among the main planetary frequencies
shown in Figure \ref{fig6}A. \citet{Scafetta2010} argued that this
oscillation is likely linked to a combination of the 8.85-year lunar
apsidal line rotation period, the first harmonic of the 9-year Saros
eclipse cycle and the 9.3-year first harmonic of the soli-lunar nodal
cycle \citep[supplement]{Cionco,Scafetta2012d}. These three lunar
cycles induce oceanic tides with an average period of about 9.1 years
\citep{Wood,Keeling} that could affect the climate system by modulating
the atmospheric and oceanic circulation.

The 10.4-year temperature cycle is variable and appears to be the
signature of the 11-year solar cycle that varies between the Jupiter-Saturn
spring tidal cycle (9.93 years) and the orbital period of Jupiter
(11.86 years). Note that in Figure \ref{fig6}B, the frequency of
this temperature signal increased in time from 1900 to 2000. This
agrees with the solar cycle being slightly longer (and smaller) at
the beginning of the 20th century and shorter (and larger) at its
end (see Figure \ref{fig2}). We also notice that the 10.46-year period
corresponds to the orbital invariant inequality $(3,-5,3,-1)$ among
Venus, Earth, Jupiter and Saturn.

The above findings were crucial for the construction of a semi-empirical
climate model based on the several astronomically identified cycles
\citep{Scafetta2010,Scafetta2013}. The model included the 9.1-year
solar-lunar cycle, the astronomical-solar cycles at 10.5, 20, 60 and,
in addition, two longer cycles with periods of 115 years (using Eq.
\ref{eq:4.14}) and a millennial cycle here characterized by an asymmetric
981-year cycle with a minimum around 1700 (the Maunder Minimum) and
two maxima in 1080 and 2060 (using Eq. \ref{eq.4.17}). The model
was completed by adding the volcano and the anthropogenic components
deduced from the ensemble average prediction of the CMIP5 global circulation
models assuming an equilibrium climate sensitivity (ECS) of about
1.5°C that is half of that of the model average, which is about 3°C.
This operation was necessary because the identified natural oscillations
already account for at least 50\% of the warming observed from 1970
to 2000. Recently, \citet{Scafetta2021} upgraded the model by adding
some higher frequency cycles.

Figure \ref{fig6}C shows the HadCRUT4.6 global surface temperature
record \citep{Morice} against the ensemble average simulations produced
by the CMIP6 global circulation models (GCMs) using historical forcings
(1850-2014) extended with three different shared socioeconomic pathway
(SSP) scenarios (2015-2100) \citep{Eyring}. Figure \ref{fig6}D shows
the same temperature record against the proposed semi-empirical astronomical
harmonic model under the same forcing conditions. The comparison between
panels C and D shows that the semi-empirical harmonic model performs
significantly better than the classical GCMs in hindcasting the 1850-2020
temperature record. It also predicts moderate warming for the future
decades, as explained in detail by \citet{Scafetta2013,Scafetta2021}.

\section{Possible physical mechanisms}

Many authors suggest that solar cycles revealed in sunspot and cosmogenic
records could derive from a deterministic non-linear chaotic dynamo
\citep{Weiss(2016),Charbonneau(2020),Charbonneau(2022)}. However,
the assumption that solar activity is only regulated by dynamical
and stochastic processes inside the Sun has never been validated mainly
because these models have a poor hindcasting capability.

We have seen how the several main planetary harmonics and orbital
invariant inequalities tend to cluster towards specific frequencies
that characterize the observed solar activity cycles. This suggests
that the strong synchronization among the planetary orbits could be
further extended to the physical processes that are responsible for
the observed solar variability.

The physical mechanisms that could explain how the planets may directly
or indirectly influence the Sun are currently unclear. It can be conjectured
that the solar dynamo might have been synchronized to some planetary
periods under the action of harmonic forcings acting on it for several
hundred million or even billion years. In fact, as pointed out by
Huygens in the 17\textsuperscript{th} century, synchronization can
occur even if the harmonic forcing is very weak but lasts long enough
\citep{Pikovsky}.

There may be two basic types of mechanisms referred to how and where
in the Sun the planetary forcing is acting. In particular, we distinguish
between the mechanisms that interact with the outer regions of the
Sun and those that act in its interior.
\begin{enumerate}
\item Planetary tides can perturb the surface magnetic activity of the Sun,
the solar corona, and thus the solar wind. The solar wind, driven
by the rotating twisted magnetic field lines \citep{Parker,Tattersall},
can reconnect with the magnetic fields of the planets when they get
closer during conjunctions. This would modulate the solar magnetic
wind density distribution and the screening efficiency of the whole
heliosphere on the incoming cosmic rays. The effect would be a modulation
of the cosmogenic records which then also act on the cloud cover.
It is also possible that the planets can focus and modulate by gravitational
lensing the flux of interstellar and interplanetary matter -- perhaps
even of dark matter -- towards the Sun and the Earth stimulating
solar activity \citep{Bertolucci,Scafetta2020,Zioutas} and, again,
contributing to clouds formation on Earth which alters the climate.
\item Gravitational planetary tides and torques could reach the interior
of the Sun and synchronize the solar dynamo by forcing its tachocline
\citep{Abreu,Stefani2016,Stefani2019,Stefani2021} or even modulate
the nuclear activity in the core \citep{Scafetta2012b,Wolff}.
\end{enumerate}
\citet{ScafettaWillson2013b} argued that these two basic mechanisms
could well complement each other. In principle, it might also be possible
that the physical solar dynamo is characterized by a number of natural
frequencies that could resonate with the external periodic forcings
yielding some type of synchronization. Let us briefly analyze several
cases.

\subsection{Mechanisms associated with planetary alignments}

The frequencies associated with planetary alignments and, in particular,
those of the Jovian planets, were found to reproduce the main observed
cycles in solar and climatic data. \citet{Scafetta2020} showed examples
of gravitational field configurations produced by a toy-model made
of four equal masses orbiting around a 10 times more massive central
body.

The Sun could feel planetary conjunctions because at least twenty-five
out of thirty-eight largest solar flares were observed to start when
one or more planets among Mercury, Venus, Earth, and Jupiter were
either nearly above the position of the flare (within $10{^\circ}$
longitude) or on the opposite side of the Sun \citep{Hung}. For example,
\citet{Morner(2015)} showed that, on January 7 2014, a giant solar
flare of class X1.2 was emitted from the giant sunspot active region
AR1944 \citep{NASA2014a}, and that the flare pointed directly toward
the Earth when Venus, Earth and Jupiter were exactly aligned in a
triple conjunction and the planetary tidal index calculated by \citet{Scafetta2012b}
peaked at the same time.

\citet{Hung} estimated that the probability for this to happen at
random was 0.039\%, and concluded that \textit{``the force or momentum
balance (between the solar atmospheric pressure, the gravity field,
and magnetic field) on plasma in the looping magnetic field lines
in solar corona could be disturbed by tides, resulting in magnetic
field reconnection, solar flares, and solar storms.''} Comparable
results and confirmations that solar flares could be linked to planetary
alignments were recently discussed in \citet{Bertolucci} and \citet{Petrakou}.

\subsection{Mechanisms associated with the solar wobbling}

The movement of the planets and, in particular, of the Jovian ones,
are reflected in the solar wobbling. \citet{Charvatova2000} and \citet{Charvatova2013}
showed that the solar wobbling around the center of mass of the solar
system forms two kinds of complex trajectories: an ordered one, where
the orbits appear more symmetric and circular, and a disordered type,
where the orbits appear more eccentric and randomly distributed. These
authors found that the alternation between these two states presents
periodicities related, for example, to the Jose ($\sim$178 years)
and Bray--Hallstatt ($\sim$2300 years) cycles.

\begin{figure*}[!t]
\centering{}\includegraphics[width=1\textwidth]{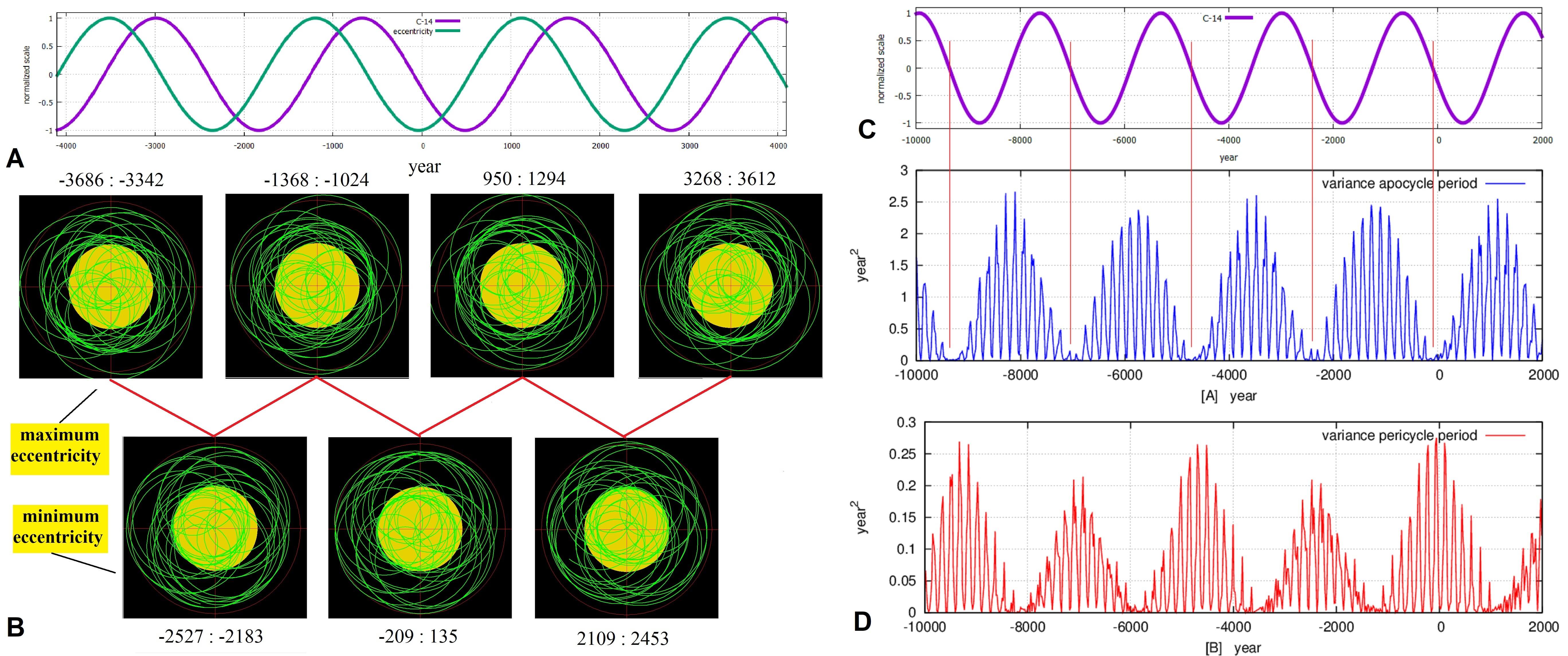}\caption{{[}A{]} The Hallstatt oscillation (2318 years) found in $\varDelta^{14}C$
(\textperthousand ) record and in the eccentricity function of the
barycenter of the planets relative to the Sun. {[}B{]} Ordered and
disordered orbits of the barycenter of the planets relative to the
Sun. {[}C and D{]} The Hallstatt oscillation found in $\varDelta^{14}C$
(\textperthousand ) record and in the apocycles and pericycles of
the orbits of the center of mass of the planets relative to the Sun.
\citep[cf.][]{Scafettaetal2016}.}
\label{fig7}
\end{figure*}

Figure \ref{fig7}A compares the Bray--Hallstatt cycle found in the
$\Delta^{14}C$ (\textperthousand) record (black) throughout the
Holocene \citep[IntCal04.14c]{Reimer} with two orbital records representing
the periods of the pericycle and apocycle orbital arcs of the solar
trajectories as extensively discussed by \citet{Scafettaetal2016}.
Figure \ref{fig7}B shows the solar wobbling for about 6000 years
where the alternation of ordered and disordered orbital patterns typically
occurs according to the Bray--Hallstatt cycle of 2318 years \citep{Scafettaetal2016}.

In particular, the astronomical records show that the Jose cycle is
modulated by the Bray--Hallstatt cycle. Figures \ref{fig7}C and
D show examples of how planetary configurations can reproduce the
Bray--Hallstatt cycle: see details in \citet{Scafettaetal2016}.
The fast oscillations correspond to the orbital invariant inequalities
with periods of 159, 171.4 and 185 years while the long beat oscillation
corresponds to the orbital invariant inequality with a period of 2318
years, which perfectly fits the Bray--Hallstatt cycle as estimated
in \citet{McCracken2013} (see Table \ref{tab5.2}). It is possible
that the pulsing dynamics of the heliosphere can periodically modulate
the solar wind termination shock layer and, therefore, the incoming
interstellar dust and cosmic ray fluxes.

\subsection{Mechanisms associated with planetary tides and tidal torques}

Discussing the tidal interactions between early-type binaries, \citet{Goldreich1989}
demonstrated that the tidal action and torques can produce important
effects in the thin overshooting region between the radiative and
the convective zone, which is very close to the tachocline. This would
translate both in tidal torques and in the onset of g-waves moving
throughout the radiative region. A similar mechanism should also take
place in late-type stars like the Sun \citep{Goodman}.

\citet{Abreu} found an excellent agreement between the long-term
solar cycles and the periodicities in the planetary tidal torques.
These authors assumed that the solar interior is characterized by
a non-spherical tachocline. Under such a condition, the planetary
gravitational forces exert a torque on the tachocline itself that
would then vary with the distribution of the planets around the Sun.
These authors showed that the torque function is characterized by
some specific planetary frequencies that match those observed in cosmogenic
radionuclide proxies of solar activity. The authors highlighted spectral
coherence at the following periods: 88, 104, 150, 208 and 506 years.
The first four periods were discussed above using alternative planetary
functions; the last period could be a harmonic of the millennial solar
cycle also discussed above and found in the same solar record \citep{Scafetta2012a,Scafetta2014b}.

\citet{Abreu} observed that the tachocline approximately coincides
with the layer at the bottom of the convection zone where the storage
and amplification of the magnetic flux tubes occur. These are the
flux tubes that eventually erupt at the solar photosphere to form
active regions. The tachocline layer is in a critical state because
it is very sensitive to small perturbations being between the radiative
zone characterized by stable stratification ($\delta<0$) and the
convective zone characterized by unstable stratification ($\delta>0$).
The proposed hypothesis is that the planetary tides could influence
the magnetic storage capacity of the tachocline region by modifying
its entropy stratification and the superadiabaticity parameter $\delta$,
thereby altering the maximum field strength of the magnetic flux tubes
that regulate the solar dynamo.

However, \citet{Abreu} also acknowledged that their hypothesis could
not explain how the tiny tidal modification of the entropy stratification
could produce an observable effect although they conjectured the presence
of a resonance mediated by gravity waves.

\begin{table*}[!t]
\begin{centering}
\begin{tabular}{cccccccc}
\hline 
 & mass & semi-major & perihelion & aphelion & mean tidal & diff. tidal & Sun rot.\tabularnewline
 & (kg) & axis (m) & (m) & (m) & elong. (m) & elong. (m) & (days)\tabularnewline
\hline 
Me & 3.30E23 & 5.79E10 & 4.60E10 & 6.98E10 & 3.0E-4 (7.5E-4) & 4.3E-4 (1.1E-3) & 37.92\tabularnewline
\hline 
Ve & 4.87E24 & 1.08E11 & 1.08E11 & 1.09E11 & 6.8E-4 (1.7E-3) & 2.6E-5 (6.6E-5) & 30.04\tabularnewline
\hline 
Ea & 5.97E24 & 1.50E11 & 1.47E11 & 1.52E11 & 3.2E-4 (7.9E-4) & 3.2E-5 (7.9E-5) & 28.57\tabularnewline
\hline 
Ma & 6.42E23 & 2.28E11 & 2.07E11 & 2.49E11 & 9.6E-6 (2.4E-5) & 5.5E-6 (1.4E-5) & 27.56\tabularnewline
\hline 
Ju & 1.90E27 & 7.79E11 & 7.41E11 & 8.17E11 & 7.1E-4 (1.8E-3) & 2.1E-4 (5.2E-4) & 26.66\tabularnewline
\hline 
Sa & 5.69E26 & 1.43E12 & 1.35E12 & 1.51E12 & 3.4E-5 (8.5E-5) & 1.2E-5 (2.9E-5) & 26.57\tabularnewline
\hline 
Ur & 8.68E25 & 2.88E12 & 2.75E12 & 3.00E12 & 6.4E-7 (1.6E-6) & 1.7E-7 (4.3E-7) & 26.52\tabularnewline
\hline 
Ne & 1.02E26 & 4.50E12 & 4.45E12 & 4.55E12 & 2.0E-7 (5.0E-7) & 1.3E-8 (3.3E-8) & 26.51\tabularnewline
\hline 
\end{tabular}
\par\end{centering}
\caption{Mean tidal elongations at the solar surface produced by all planets.
\textquotedblleft Diff. tidal elongation\textquotedblright{} is the
difference between the tides at perihelion and aphelion. The 26.5-day
mean solar rotation as seenby the planets. Tidal elongations are calculated
for Love-number 3/2 and 15/4, the latter being inside parentheses.
\citep[cf.][]{Scafetta2012b}.}
\label{tab6.1}
\end{table*}

The planetary tidal influence on the solar dynamo has been rather
controversial because the tidal accelerations at the tachocline layer
are about 1000 times smaller than the accelerations of the convective
cells \citep{Jager}. \citet{Scafetta2012b} calculated that the gravitational
tidal amplitudes produced by all the planets on the solar chromosphere
are of the order of one millimeter or smaller (see Table \ref{tab6.1}).
More recently, \citet{Charbonneau(2022)} critiqued \citet{Stefani2019,Stefani2021}
by observing that also the planetary tidal forcings of Jupiter and
Venus could only exert a ``homeopathic'' influence on the solar
tachocline concluding that they should be unable to synchronize the
dynamo. \citet{Charbonneau(2022)} also observed that even angular
momentum transport by convective overshoot into the tachocline would
be inefficient and concluded that synchronization could only be readily
achieved in presence of high forcing amplitudes, stressing the critical
need for a powerful amplification mechanism.

While it is certainly true that the precise underlying mechanism is
not completely understood, the rough energetic estimate that 1 mm
tidal height corresponds to 1 m/s velocity at the tachocline level
might still entail sufficient capacity for synchronization by changing
the (sensitive) field storage capacity \citep{Abreu} or by synchronizing
that part of $\alpha$ that is connected with the Tayler instability
or by the onset of magneto-Rossby waves at the tachocline \citep{Dikpati2017,Zaqarashvili(2018)}.
In all cases, it could be possible that only a few high-frequency
planetary forcing (e.g. the 11.07-year Venus-Earth-Jupiter tidal model)
are able to efficiently synchronize the solar dynamo \citep{Stefani2016,Stefani2018,Stefani2019}.
At the same time, additional and longer solar cycles could emerge
when some feature of the dynamo is also modulated by the angular momentum
exchange associated with the solar wobbling \citep{Stefani2021}.
Finally, \citet{Albert} proposed that stochastic resonance could
explain the multi-secular variability of the Schwabe cycle by letting
the dynamo switch between two distinct operating modes as the solution
moves back and forth from the attraction basin of one to the other.

Alternatively, the problem of the tidal ``homeopathic'' influence
on the tachocline could be solved by observing that tides could play
some more observable role in the large solar corona where the solar
wind originates, or in the wind itself at larger distances from the
Sun where the tides are stronger, or even in the solar core where
they could actually trigger a powerful response from nuclear fusion
processes. Let us now discuss the latter hypothesis.

\subsection{A possible solar amplification of the planetary tidal forcing}

A possible amplification mechanism of the effects of the tidal forcing
was introduced by \citet{Wolff} and \citet{Scafetta2012b}.

\citet{Wolff} proposed that tidal forcing could act inside the solar
core inducing waves in the plasma by mixing the material and carrying
fresh fuel to the deeper and hotter regions. This mechanism would
make solar-type stars with a planetary system slightly brighter because
their fuel would burn more quickly.

\citet{Scafetta2012b} further developed this approach and introduced
a physical mechanism inspired by the mass-luminosity relation of main-sequence
stars. The basic idea is that the luminosity of the core of the Sun
can be written as

\begin{equation}
L(t)\approx L_{\varodot}+A\cdot\dot{\Omega}_{tidal}(t),\label{eq:6.2-1}
\end{equation}
where $L_{\varodot}$ is the baseline luminosity of the star without
planets and $\Delta L_{tidal}(t)=A\cdot\dot{\Omega}_{tidal}(t)$ is
the small luminosity increase induced by planetary tides inside the
Sun. $\dot{\Omega}_{tidal}(t)$ is the rate of the gravitational tidal
energy which is continuously dissipated in the core and $A$ is the
amplification factor related to the triggered luminosity production
via H-burning.

To calculate the magnitude of the amplification factor $A$ we start
by considering the Hertzsprung-Russell \emph{mass-luminosity relation},
which establishes that, if the mass of a star increases, its luminosity
$L$ increases as well. In the case of a G-type main-sequence star,
with luminosity $L$ and mass $M=M_{\varodot}+\Delta M$, the mass-luminosity
relation approximately gives

\begin{equation}
\frac{L}{L_{\varodot}}\approx\left(\frac{M}{M_{\varodot}}\right)^{4}\approx1+\frac{4\Delta M}{M_{\varodot}},\label{eq:6.1}
\end{equation}
where $L_{\varodot}$ is the solar luminosity and $M_{\varodot}$
is the mass of the Sun \citep{Duric}. By relating the luminosity
of a star to its mass, the Hertzsprung-Russell relation suggests a
link between the luminosity and the gravitational power continuously
dissipated inside the star.

The total solar luminosity is

\begin{equation}
L_{\varodot}=4\pi(1AU)^{2}\times TSI=3.827\cdot10^{26}~W~,\label{eq:6.3}
\end{equation}
where 1 AU = $1.496\cdot10^{11}$ m is the average Sun-Earth distance,
and TSI is the total solar irradiance 1360.94 $W/m^{2}$ at 1 AU.
Every second, the core of the Sun transforms into luminosity a certain
amount of mass according to the Einstein equation $E=mc^{2}$. If
$dL(r)$ is the luminosity produced inside the shell between $r$
and $r+dr$ \citep{Bahcall01,Bahcall}, the mass transformed into
light every second in the shell is

\begin{equation}
\frac{d\dot{m}(r)}{dr}=\frac{1}{c^{2}}~\frac{dL(r)}{dr},\label{eq:6.4}
\end{equation}
where $c=2.998\cdot10^{8}~m/s$ is the speed of the light and $r$
is the distance from the center of the Sun.

The transformed material can be associated with a correspondent loss
of gravitational energy of the star per time unit $\dot{U}_{\varodot}$,
which can be calculated using Eq. \ref{eq:6.4} as

\begin{align}
\dot{U}_{\varodot} & =\frac{1}{2}~G\int_{0}^{R_{S}}m_{\varodot}(r)~\frac{d\dot{m}(r)}{dr}~\frac{1}{r}~dr\label{eq:6.5}\\
= & \frac{1}{2}~\frac{G}{c^{2}}\int_{0}^{R_{S}}m_{\varodot}(r)~\frac{dL(r)}{dr}~\frac{1}{r}~dr=3.6\cdot10^{20}~W\nonumber 
\end{align}
where the initial factor $1/2$ is due to the virial theorem, $m_{\varodot}(r)$
is the solar mass within the radius $r\leq R_{S}$ and $L(r)$ is
the luminosity profile function derived by the standard solar model
\citep{Bahcall01,Bahcall}.

The gravitational forces will do the work necessary to compensate
for such a loss of energy to restore the conditions for the H-burning.
In fact, the solar luminosity would decrease if the Sun's gravity
did not fill the vacuum created by the H-burning, which reduces the
number of particles by four ($4H\rightarrow1He$). At the same time,
the nucleus of He slowly sinks releasing additional potential energy.
All this corresponds to a gravitational work in the core per time
unit, $\dot{\Omega}_{\varodot}$, that is associated with light production.

The basic analogy made by \citet{Scafetta2012b} is that $\dot{\Omega}_{\odot}$
should be of the same order of magnitude as the rate of the gravitational
energy loss due to H-fusion ($\dot{\Omega}_{\varodot}\approx\dot{U}_{\varodot}$).
Moreover, the energy produced by the dissipation of the tidal forces
in the core should be indistinguishable from the energy produced by
the other gravitational forces in the Sun. Thus, it is as if tides
added some gravitational power that becomes $\dot{\Omega}_{\odot}+\dot{\Omega}_{tidal}$.

For small perturbations, since light production is directly related
both to the solar mass and to the gravitational power dissipated inside
the core, \citet{Scafetta2012b} assumed the equivalence

\begin{equation}
\frac{\Delta M}{M_{\varodot}}\equiv\frac{\dot{\Omega}_{tidal}}{\dot{\Omega}_{\varodot}},
\end{equation}
where $\dot{\Omega}_{tidal}$ is the tidal perturbing power dissipated
inside the Sun and $\dot{\Omega}_{\odot}\equiv\dot{U}_{\varodot}$
is the rate of the gravitational energy lost by the Sun through H-burning.
Thus, from Eqs. \ref{eq:6.2-1} and \ref{eq:6.1} we get

\begin{equation}
L(t)\approx L_{\varodot}+\frac{4L_{\varodot}}{\dot{\Omega}_{\varodot}}\dot{\Omega}_{tidal}(t)=L_{\varodot}+A\cdot\dot{\Omega}_{tidal}(t),\label{eq:6.2}
\end{equation}
where the amplification factor is

\begin{equation}
A=4\frac{L_{\varodot}}{\dot{\Omega}_{\varodot}}\approx4\frac{L_{\varodot}}{\dot{U}_{\varodot}}\approx4.25\cdot10^{6}.\label{eq:6.6}
\end{equation}
Eq. \ref{eq:6.6} means that any little amount of gravitational power
dissipated in the core (like that induced by planetary tidal forcing)
could be amplified by a factor of the order of one million by nuclear
fusion. This could be equivalent to having gravitational tidal amplitudes
amplified from 1 mm to 1 kilometer at the tachocline. This amplification
could solve the problem of the ``homeopathic'' gravitational tidal
energy contribution highlighted by \citet{Charbonneau(2022)}.

\begin{figure*}[!t]
\begin{centering}
\includegraphics[width=1\textwidth]{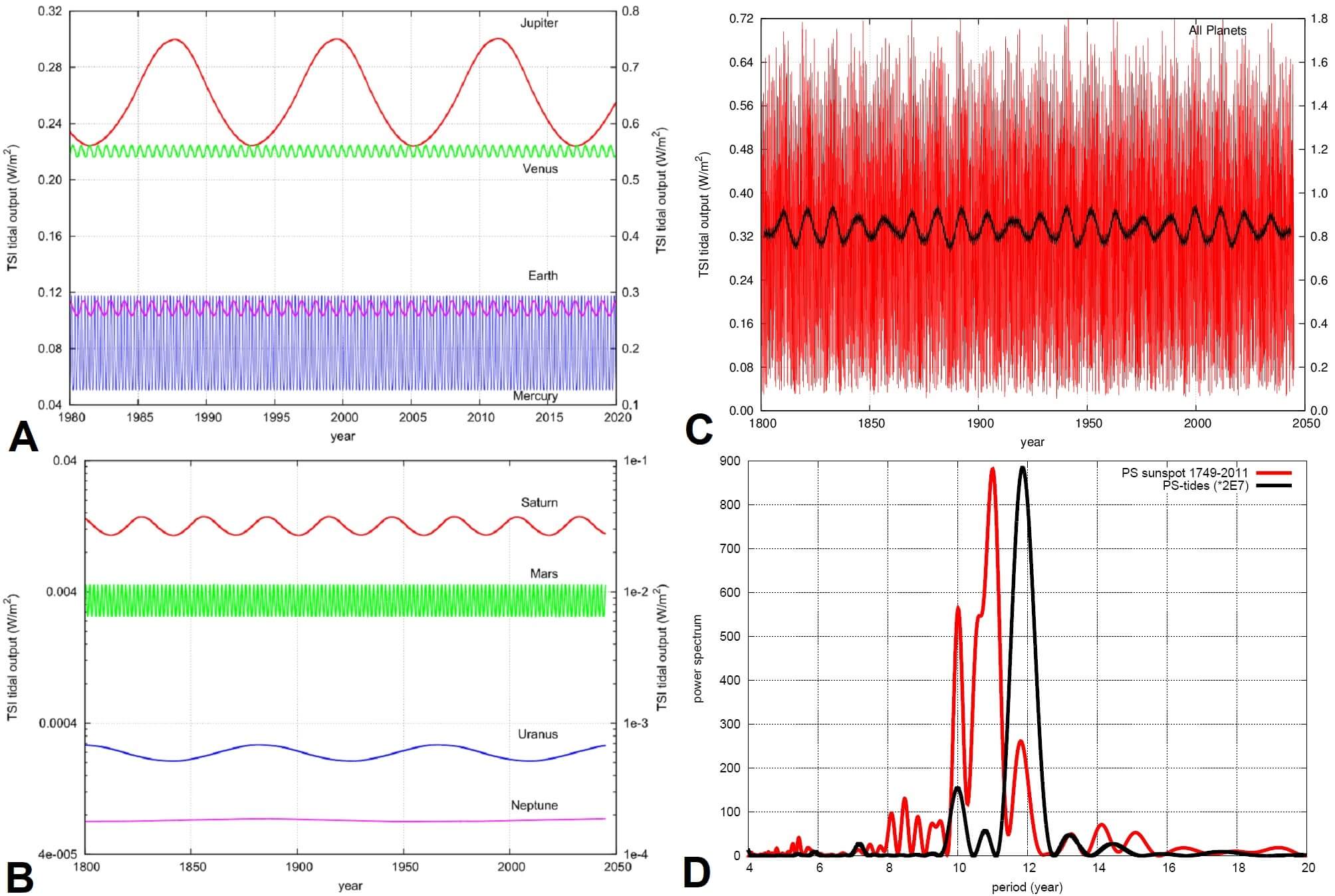}
\par\end{centering}
\caption{{[}A-B{]} Theoretical TSI enhancement induced by the tides of each
planet on the Sun obtained using \citet{Scafetta2012b} amplification
hypothesis; the Love numbers are 3/2 (left axis) and 15/4 (right axis).
{[}C{]} The same as produced by the tides of all the planets. {[}D{]}
Lomb-periodogram spectral analysis of the sunspot number record (red)
and of the tidal function (black) produced by all the planets.}
\label{fig8}
\end{figure*}

By using such a large amplification factor the estimated gravitational
power $\dot{\Omega}_{tidal}$ dissipated inside the solar core, \citet{Scafetta2012b}
calculated the tidally-induced TSI produced by each of the planets
(Figure \ref{fig8}A and B), as well as that of all the planets together
(Figure \ref{fig8}C). The sequence of the relative tidal relevance
of the planets is Jupiter, Venus, Earth, Mercury, Saturn, Mars, Uranus
and Neptune. The mean enhancement of their overall tidally-induced
TSI is of the order of 0.3-0.8 $W/m^{2}$, depending on the specific
Love number of the tides (see Figure \ref{fig8}C). However, on shorter
time scales the tides could produce TSI fluctuations up to 0.6-1.6
$W/m^{2}$ in absence of dampening mechanisms. In particular, on a
decadal time scale, the TSI fluctuations due to Jupiter and Saturn
could reach amplitudes of 0.08-0.20 $W/m^{2}$ (see the black curve
in Figure \ref{fig8}C).

If the luminosity flux reaching the tachocline from the radiative
zone is modulated by the contribution of tidally-induced luminosity
oscillations with a TSI amplitude of the order of 0.01-0.10 $W/m^{2}$,
the perturbation could be sufficiently energetic to tune the solar
dynamo with the planetary frequencies. The dynamo would then further
amplify the luminosity signal received at the tachocline up to $\sim1$
$W/m^{2}$ amplitudes as observed in TSI cycles \citep{Willson2003}.

Figure \ref{fig8}D compares the periodograms of the sunspot number
record and of the planetary luminosity signal shown in Figure \ref{fig8}C.
The two side frequency peaks at about 10 years (J/S-spring tide) and
11.86 years (J-tide) perfectly coincide in the two spectral analyses.
The central frequency peak at about 10.87 years shown only by sunspot
numbers could be directly generated by the solar dynamo excited by
the two tidal frequencies \citep{Scafetta2012a} or other mechanisms
connected with the dynamo as discussed above.

An obvious objection to the above approach is that the Kelvin-Helmholtz
time-scale \citep{Mitalas,Stix} predicts that the light journey from
the core to the convective zone requires $10^{4}$ to $10^{8}$ years.
Therefore, the luminosity fluctuations produced inside the core could
be hardly detectable because they would be smeared out before reaching
the convective zone. At most, there could exist only a slightly enhanced
solar luminosity related to the overall tidally-induced TSI mean enhancement
of the order of 0.3-0.8 $W/m^{2}$ as shown in Figure \ref{fig8}C.

However, several different mechanisms may be at work. In fact, the
harmonic tidal forcing acts simultaneously throughout the core and
in the radiative zone , and simultaneously produces everywhere a synchronized
energy oscillation that can be amplified in the core as discussed
above. This would give rise to modulated seismic waves \textit{\emph{(g
and p-mode oscillations) that can}} propagate from the inner core
up to the tachocline region in a few hours because the sound speed
inside the Sun is a few hundred kilometers per second \citep{Hartlep,Ahuir,Barker}.
These waves might also couple with the g-waves produced in the tachocline
\citep{Goodman} producing a in the tachocline region sufficiently
strong to synchronize the solar dynamo with the planetary tidal frequencies.

\section{Conclusion}

Many empirical evidences suggest that planetary systems can self-organize
in synchronized structures although some of the physical mechanisms
involved are still debated.

We have shown that the high synchronization of our own planetary system
is nicely revealed by the fact that the ratios of the orbital radii
of adjacent planets, when raised to the 2/3rd power, express the simple
ratios found in harmonic musical consonances while those of the mirrored
ones follow the simple, elegant, and highly precise scaling-mirror
symmetry Eq. \ref{eq:1.0} \citep{Bank}.

The solar system is made of synchronized coupled oscillators because
it is characterized by a set of frequencies that are linked to each
other by the harmonic Eq. \ref{eq:1.1}, which are easily detected
in the solar wobbling. Thus, it is then reasonable to hypothesize
that the solar activity could be also tuned to planetary frequencies.

We corroborated this hypothesis by reviewing the many planetary harmonics
and orbital invariant inequalities that characterize the planetary
motions and observing that often their frequencies correspond to those
of solar variability.

It may be objected that, since the identified planetary frequencies
are so numerous, it could be easy to occasionally find that some of
them roughly correspond to those of the solar cycles. However, the
fact is that the planetary frequencies of the solar system, from the
monthly to the millennial time scales, are not randomly distributed
but tend to cluster around some specific values that quite well match
 those of the main solar activity.

Thus, it is rather unlikely that the results shown in Figures \ref{fig2}-\ref{fig6}
are just occasional. In some cases, our proposed planetary models
have even been able to predict the time-phase of the solar oscillations
like that of the Schwabe 11-year sunspot cycle throughout the last
three centuries, as well as those of the secular and millennial modulations
throughout the Holocene. The two main planetary models that could
explain the Schwabe 11-year cycle and its secular and millennial variation
involve the planets Venus, Earth, Jupiter and Saturn, as it was initially
suggested by \citet{Wolf}. We further suggest that the Venus-Earth-Jupiter
model and the Jupiter-Saturn model could be working complementary
to each other.

The alternative hypothesis that the solar activity is regulated by
an unforced internal dynamics alone (i.e. by an externally unperturbed
solar dynamo) has never been able to reproduce the variety of the
observed oscillations. In fact, standard MHD dynamo models are not
self-consistent and do not directly explain the well-known 11-year
solar cycle nor they are able to predict its timing without assuming
a number of calibrated parameters \citep{Jiang,Tobias2002}.

There have been several objections to a planetary theory of solar
variability. For example, \citet{Smythe} claimed that planetary cycles
and conjunctions could not predict the timing of grand solar minima,
like the Maunder Minimum of the 17\textsuperscript{th} century. However,
\citet{Scafetta2012a}  developed a solar-planetary model able to
predict all the grand solar maxima and minima of the last millennium
(Figure \ref{fig4}).

Other authors reasonably claimed that planetary gravitational tides
are too weak to modulate solar activity \citep{Charbonneau,Jager,Charbonneau(2022)};
yet,several empirical evidences support the importance of their role
\citep{Abreu,Scafetta2012b,Stefani2016,Stefani2019,Wolff}. \citet{Stefani2016,Stefani2021}
proposed that the Sun could be at least synchronized by the tides
of Venus, Earth and Jupiter producing an 11.07-year cycle that reasonably
fits the Schwabe cycle. Longer cycles could be produced by a dynamo
excited by angular momentum transfer from Jupiter and Saturn. Instead,
\citet{Scafetta2012b} proposed that, in the solar core, the effects
of the weak tidal forces could be amplified one million times or more
due to an induced increase in the H-burning, thus providing a sufficiently
strong forcing to synchronize and modulate the solar dynamo with planetary
harmonics at multiple time scales.

Objections to the latter hypothesis, based on the slow light propagation
inside the radiative zone according to the Kelvin--Helmholtz timescale
\citep{Mitalas,Stix}, could be probably solved. In fact, tidal forces
are believed to favor the onset of g-waves moving back and forth throughout
the radiative region of the Sun \citep{Ahuir,Barker}. Thus, g-waves
themselves could be amplified and modulated in the core by the tidally
induced H-burning enhancement \citep{Scafetta2012b}. Then, both tidal
torques and g-waves could cyclically affect the tachocline region
at the bottom of the convective zone and synchronize the solar dynamo.

Alternatively, planetary alignments can also modify the large-scale
electromagnetic and gravitational structure of the planetary system
altering the space weather in the solar system. For example, in coincidence
of planetary alignments, an increase of solar flares has been observed
\citep{Hung,Bertolucci,Petrakou}. The solar wobbling, which reflects
the motion of the barycenter of the planets, changing from more regular
to more chaotic trajectories, correlates well with some long climate
cycles like the Bray-Hallstatt cycle (2100-2500 years) \citep{Charvatova2000,Charvatova2013,Scafettaetal2016}.
Finally, \citet{Scafettaetal2020} showed that the infalling meteorite
flux on the Earth presents a 60-year oscillation coherent with the
variation of the eccentricity of Jupiter\textquoteright s orbit induced
by Saturn. The falling flux of meteorites and interplanetary dust
would then contribute to modulate cloud formation.

In conclusion, much empirical evidence suggests that planetary oscillations
should be able to modulate the solar activity and even the Earth\textquoteright s
climate, although several open physical issues remain open. These
results stress the importance of identifying the relevant planetary
harmonics, the solar activity cycles and the climate oscillations
as phenomena that, in many cases, are interconnected. This approach
could be useful to predict both solar and climate variability using
harmonic constituent models as it is currently done for oceanic tides.
We think that the theory of a planetary modulation of solar activity
should be further developed because no clear alternative theory exists
to date capable to explain the observed planetary-solar interconnected
periodicities.

\subsection*{Author Contributions}

NS wrote the initial draft and prepared the figures. NS and AB contributed
to the discussion and the revision of the submitted version.

\subsection*{Conflict of Interest}

The authors declare that the research was conducted in the absence
of any commercial or financial relationships that could be construed
as a potential conflict of interest.

\onecolumn


\begin{thebibliography}{999}
\bibitem[Abreu et al.(2012)]{Abreu} Abreu, J. A., Beer, J., Ferriz-Mas,
A., McCracken, K. G., and Steinhilber, F. (2012). Is there a planetary
influence on solar activity? \textit{Astron. Astrophys.} 548, A88.
doi:10.1051/0004-6361/201219997

\bibitem[Albert et al.(2021)]{Albert}Albert, C., Ferriz-Mas, A.,
Gaia, F., and Ulzega, S. (2021). Can Stochastic Resonance Explain
Recurrence of Grand Minima? ApJL 916, L9.

\bibitem[Agnihotri and Dutta(2003)]{Agnihotri}Agnihotri, R., and
Dutta, K. (2003). Centennial scale variations in monsoonal rainfall
(Indian, east equatorial and Chinese monsoons): manifestations of
solar variability. \emph{Current Science} 85, 459--463.

\bibitem[Agol et al.(2021)]{Agol}Agol, E., Dorn, C., Grimm, S. L.,
Turbet, M., Ducrot, E., Delrez, L., et al. (2021). Refining the Transit-timing
and Photometric Analysis of TRAPPIST-1: Masses, Radii, Densities,
Dynamics, and Ephemerides. \emph{Planetary Science Journal} \textbf{2},
1.

\bibitem[Ahuir et al.(2021)]{Ahuir}Ahuir, J., Mathis, S., Amard L.
2021. Dynamical tide in stellar radiative zones. \emph{Astronomy \&
Astrophysics} \textbf{651}, A3.

\bibitem[Aschwanden(2018)]{Aschwanden}Aschwanden, M. J. (2018). Self-organizing
systems in planetary physics: Harmonic resonances of planet and moon
orbits. \emph{New Astronomy} 58, 107--123.

\bibitem[Bahcall et al.(2001)]{Bahcall01} Bahcall J. N., Pinsonneault
M. H., and Basu S. (2001). Solar models: current epoch and time dependences,
neutrinos, and helioseismological properties. \emph{Astrophysical
Journal} 555, 990--1012.

\bibitem[Bahcall et al.(2005)]{Bahcall} Bahcall J. N., Serenelli
A., and Basu S. (2005). New solar opacities, abundances, helioseismology,
and neutrino fluxes. \emph{Astrophysical Journal} 621, L85--L88.

\bibitem[Bank and Scafetta(2022)]{Bank}Bank, M. J., and Scafetta,
N. (2022). Scaling, Mirror Symmetries and Musical Consonances Among
the Distances of the Planets of the Solar System. \emph{Front. Astron.
Space Sci.} 8:758184. doi: 10.3389/fspas.2021.758184

\bibitem[Bard et al.(2000)]{Bard} Bard, E., Raisbeck, G., Yiou, F.,
and Jouzel J. (2000). Solar irradiance during the last 1200 years
based on cosmogenic nuclides. \emph{Tellus} 52B, 985--992.

\bibitem[Barker and Ogilvie(2010)]{Barker}Barker, A. J., Ogilvie,
G.I. (2010). On internal wave breaking and tidal dissipation near
the centre of a solar-type star. Monthly Notices of the Royal Astronomical
Society 404, 1849--1868.

\bibitem[Bartels(1934)]{Bartels} Bartels, J. (1934). Twenty-seven
day recurrences in terrestrial-magnetic and solar activity, 1923--1933.
\emph{J. Geophysical Research} 39, 201-202a, doi: 10.1029/TE039i003p00201.

\bibitem[Battistini(2011)]{Battistini}Battistini, A. (2011). Il ciclo
undecennale del sole secondo Bendandi (The 11-year solar cycle according
to Bendandi). \emph{New Ice Age}, \href{http://daltonsminima.altervista.org/?p=8669}{http://daltonsminima.altervista.org/?p=8669}.

\bibitem[Beck(2000)]{Beck2000}Beck, J. (2000). A comparison of differential
rotation measurements. Solar Physics. 191: 47--70.

\bibitem[Beer et al.(2018)]{Beer}Beer, J., Tobias, S.M., Weiss, N.O.
(2018). On long-term modulation of the Sun\textquoteright s magnetic
cycle. \emph{Mon. Not. R. Astron. Soc.} 473, 1596--1602.

\bibitem[Bendandi(1931)]{Bendandi} Bendandi, R. (1931). \emph{Un
principio fondamentale dell\textquoteright Universo} (A fundamental
principle of the Universe), Faenza, Osservatorio Bendandi.

\bibitem[Bertolucci et al.(2017)]{Bertolucci} Bertolucci, S., Zioutas,
K., Hofmann, S., and Maroudas, M. (2017). The Sun and its planets
as detectors for invisible matter. \textit{Physics of the Dark Universe}
17, 13. doi:10.1016/j.dark.2017.06.001

\bibitem[Bigg(1967)]{Bigg}Bigg E.K. (1967). Influence of the planets
Mercury on Sunspots. The Astronomical Journal 72, 463--466.

\bibitem[Bollinger(1952)]{Bollinger}Bollinger, C. J. (1952). A 44.77
year Jupiter-Earth-Venus configuration Sun-tide period in solar-climate
cycles. \emph{Academy of Science for 1952 -- Proceedings of the Oklahoma},
307--311. (\href{http://digital.library.okstate.edu/oas/oas_pdf/v33/v307_311.pdf}{http://digital.library.okstate.edu/oas/oas\_pdf/v33/v307\_311.pdf})

\bibitem[Brown(1900)]{Brown} Brown, E. W. (1900). A Possible Explanation
of the Sun-spot Period. \emph{Monthly Notices of the Royal Astronomical
Society} 60, 599--606.

\bibitem[Bucha et al.(1985)]{Bucha} Bucha, V., Jakubcov\'a, I.,
and Pick, M. (1985). Resonance frequencies in the Sun\textquoteright s
motion. \emph{Studia Geophysica et Geodaetica} 29, 107--111.

\bibitem[Charbonneau(2002)]{Charbonneau}Charbonneau, P. (2002). The
rise and fall of the first solar cycle model. \emph{Journal for the
History of Astronomy} 33(4), 351--372.

\bibitem[Charbonneau(2020)]{Charbonneau(2020)}Charbonneau P. (2020).
Dynamo Models of the Solar Cycle. \emph{Living Rev. Sol. Phys.} 17,
4.

\bibitem[Charbonneau(2022)]{Charbonneau(2022)}Charbonneau P (2022)
External Forcing of the Solar Dynamo. \emph{Front. Astron. Space Sci.}
9:853676.

\bibitem[Cauquoin et al.(2014)]{Cauquoin} Cauquoin, A., Raisbeck,
G. M., Jouzel, J., Bard, E., and ASTER Team (2014). No evidence for
planetary influence on solar activity 330000 years ago. \emph{Astron.
Astrophys.} 561, A132.

\bibitem[Chang and Yu(1981)]{ChangYu1981}Chang, S., and Yu, Z. (1981).
Historical records of meteorite falls in China and their time series
analysis. \emph{National Institute of Polar Research}, Memoirs, Special
Issue 20, 276--284, 1981.

\bibitem[Charv{\'a}tov{\'a}(2000)]{Charvatova2000}Charv\'atov\'a,
I. (2000). Can origin of the 2400-year cycle of solar activity be
caused by solar inertial motion? \emph{Ann. Geophys.} 18, 399--405.
doi:10.1007/s00585-000-0399-x, 2000.

\bibitem[Charv\'atov\'a and Hejda(2014)]{Charvatova2013}Charv\'atov\'a,
I., and Hejda, P. (2014). Responses of the basic cycle of 178.7 years
and 2402 years in solar-terrestrial phenomena during Holocene. \emph{Pattern
Recogn. Phys.}, 2, 21--26. doi:10.5194/PRP-2-21-2014

\bibitem[Cionco and Pavlov(2018)]{Cionco2018}Cionco, R. G., Pavlov,
D. A. (2018). Solar barycentric dynamics from a new solar-planetary
ephemeris. Astron. Astrophys. 615, A153.

\bibitem[Cionco et al.(2021)]{Cionco}Cionco, R. G., Kudryavtsev,
S. M., Soon, W. W.-H. (2021). Possible origin of some periodicities
detected in solar-terrestrial studies: Earth's orbital movements.
\emph{Earth and Space Science} 8, e2021EA001805.

\bibitem[Cole and Bushby(2014)]{Cole}Cole, L. C., Bushby, P. J. (2014).
Modulated cycles in an illustrative solar dynamo model with competing
\textgreek{a}-effects. \emph{Astronomy \& Astrophysics} 563, A116.

\bibitem[de Vries(1958)]{de Vries}de Vries, H. (1958). Variations
in concentration of radiocarbon with time and location on Earth. \emph{Proc.
K. Ned. Akad. Wet. B}, 61, 94--102.

\bibitem[Dicke(1978)]{Dicke(1978)}Dicke, R.H. (1978). Is there a
chronometer hidden deep in the Sun? \emph{Nature} 276, 676.

\bibitem[Dikpati and Gilman(2007)]{Dikpati}Dikpati, M., Gilman, P.
A. (2007). Global solar dynamo models: simulations and predictions
of cyclic photospheric fields and long-term non-reversing interior
fields. \emph{New J. Phys.} 9, 297.

\bibitem[Dikpati et al.(2017)]{Dikpati2017}Dikpati, M., Cally, P.S.,
McIntosh, S.W., Heifetz, E.: 2017, The origin of the \textquotedblleft seasons\textquotedblright{}
in space weather. \emph{Sci. Rep.} 7, 14750.

\bibitem[Duric(2004)]{Duric} Duric N. (2004). \emph{Advanced astrophysics},
Cambridge University Press. pp. 19.

\bibitem[Eyring et al.(2016)]{Eyring}Eyring, V., Bony, S., Meehl,
G.A., \emph{et al.} (2016). Overview of the Coupled Model Intercomparison
Project Phase 6 (CMIP6) experimental design and organization. \emph{Geoscientific
Model Development}, 9 (5), 1937--1958. \href{https://doi.org/10.5194/gmd-9-1937-2016}{https://doi.org/10.5194/gmd-9-1937-2016}

\bibitem[Etz(2000)]{Etz}Etz, D. V. (2000). Conjunctions of Jupiter
and Saturn. \emph{Journal of the Royal Astronomical Society of Canada}
94, 174--178.

\bibitem[Fairbridge and Shirley(1987)]{Fairbridge}Fairbridge, R.
W., Shirley, J. H. (1987). Prolonged minima and the 179-yr cycle of
the solar inertial motions. Solar Phys. 110, 191--210.

\bibitem[Fr\"ohlich(2006)]{Frohlich} Fr\"ohlich, C. (2006). Solar
irradiance variability since 1978: revision of the PMOD composite
during solar cycle 21. \emph{Space Science Reviews} 125, 53--65.

\bibitem[Geddes and King-Hele(1983)]{Geddes}Geddes, A.B., King-Hele,
D.G. (1983). Equations for mirror symmetries among the distances of
the planets. \emph{Quarterly Journal of the Royal Astronomical Society}
\textbf{24}, 10--13.

\bibitem[Godwin(1992)]{Godwin}Godwin, J. (1992). \emph{The Harmony
of the Spheres: The Pythagorean Tradition in Music}. Inner Traditions,
Rochester, Vermont USA.

\bibitem[Goldreich and Nicholson(1989)]{Goldreich1989}Goldreich,
P., Nicholson, P. (1989). Tidal Friction in Early-Type Stars. Astrophysical
Journal 342, 1079--1084.

\bibitem[Goodman and Dickson(1998)]{Goodman}Goodman, J., Dickson,
E.S. (1998). Dynamical Tide in Solar-Type Binaries . Astrophysical
Journal 507, 938--944.

\bibitem[Gurgenashvili et al.(2022)]{Gurgenashvili} Gurgenashvili,
E., Zaqarashvili, T. V., Kukhianidze, V., Reiners, A., Reinhold, T.,
Lanza, A. F. (2022). Rieger-type cycles on the solar-like star KIC
2852336. Astronomy \& Astrophysics 660, A33.

\bibitem[Haar(1948)]{Haar}ter Haar, D. (1948). Recent theories about
the origin of the solar system. \emph{Science New Series}, 107, 405--411.

\bibitem[Hale(1908)]{Hale}Hale, G. E. (1908). On the probable existence
of a magnetic field in sun-spots. The Astrophysical Journal 28: 315.

\bibitem[Hartlep and Mansour(2005)]{Hartlep}Hartlep T., Mansour N.
N. (2005). Acoustic wave propagation in the Sun. Center for Turbulence
Research Annual Research Briefs, Stanford University, pp. 357--365.

\bibitem[Hung(2007)]{Hung} Hung, C.-C. (2007). Apparent Relations
Between Solar Activity and Solar Tides Caused by the Planets. \emph{NASA
report/TM- 2007-214817}. Available at: \href{http://ntrs.nasa.gov/search.jsp?R=20070025111}{http://ntrs.nasa.gov/search.jsp?R=20070025111}.

\bibitem[Jager and Versteegh(2005)]{Jager} de Jager C., and Versteegh
G. J. M. (2005). Do Planetary Motions Drive Solar Variability? \emph{Solar
Physics} 229, 175--179.

\bibitem[Jakubcov\'a and Pick(1986)]{Jakubcova}Jakubcov\'a, I.,
and Pick, M. (1986). The planetary system and solar-terrestrial phenomena.
\emph{Studia Geophysica et Geodaetica} 30, 224--235.

\bibitem[Jiang et al.(2007)]{Jiang}Jiang, J., Chatterjee, P., Choudhuri,
A. R. (2007). Solar activity forecast with a dynamo model. \emph{Monthly
Notices of the Royal Astronomical Society} 381, 1527--1542.

\bibitem[Jose(1965)]{Jose} Jose, P. D. (1965). Sun's motion and sunspots,
\textit{Astrophysical Journal} 70, 193. doi:10.1086/109714

\bibitem[Keeling and Whorf(2000)]{Keeling}Keeling, C. D., Whorf,
T. P. (2000). The 1,800-year oceanic tidal cycle: A possible cause
of rapid climate change. \emph{PNAS} 97 (8), 3814--3819.

\bibitem[Kepler(1606)]{Kepler}Kepler, J. (1606). \emph{De Stella
Nova in Pede Serpentarii}, Typis Pauli Sessii, Pragae.

\bibitem[Kerr(2001)]{Kerr}Kerr, R. A. (2001). A variable sun paces
millennial climate. \emph{Science} 294, 1431--1433.

\bibitem[Kopp and Lawrence(2005a)]{Kopp2005a} Kopp, G., and Lawrence,
G. (2005a). The Total Irradiance Monitor (TIM): Instrument; Design.
\emph{Solar Physics} 230, 91--109.

\bibitem[Kopp et al.(2005b)]{Kopp2005b} Kopp, G., Heuerman, K., and
Lawrence, G. (2005b). The Total Irradiance Monitor; (TIM): Instrument
Calibration. \emph{Solar Physics} 230, 111--127.

\bibitem[Kotov(2020)]{Kotov}Kotov, V.A. (2020). Rotation anomaly
of the Sun. Astron. Nachr. 341, 588-- 594.

\bibitem[Kotov and Haneychuk(2020)]{KotovH}Kotov, V.A., Haneychuk,
V.I. (2020). Oscillations of solar photosphere: 45 years of observations.
\emph{Astron. Nachr.} 341, 595-599.

\bibitem[Landscheidt(1999)]{Landscheidt(1999)}Landscheidt, T. (1999).
Extrema in sunspot cycle linked to Sun's motion. S\emph{olar Physics}
189, 415--426.

\bibitem[Ljungqvist(2010)]{Ljungqvist} Ljungqvist F. C. (2010). A
new reconstruction of temperature variability in the extra-tropical
Northern Hemisphere during the last two millennia. \emph{Geografiska
Annaler Series A} 92, 339--351.

\bibitem[Lovett(1895)]{Lovett} Lovett, E. O. (1895). The Great Inequality
of Jupiter and Saturn. \emph{Astronomical Journal}, 351, 113--127.

\bibitem[Lüdecke et al.(2015)]{L=0000FCdecke}Lüdecke, H.-J., Weiss,
C.O., Hempelmann, A. (2015). Paleoclimate forcing by the solar De
Vries/Suess cycle. \emph{Clim. Past Discuss.} 11, 279-305.

\bibitem[Macario-Rojas et al.(2018)]{Macario-Rojas}Macario-Rojas,
A., Smith, K. L., Roberts, P. C. E. (2018). Solar activity simulation
and forecast with a flux-transport dynamo. MNRAS 479, 3791--3803.

\bibitem[McCracken et al.(2001)]{McCracken2001} McCracken, K. G.,
Dreschhoff, G. A., Smart, D. F., and Shea, M. A. (2001). Solar cosmic
ray events for the period 1561-1994: 2. The Gleissberg periodicity.
\textit{Geophys. Res. Lett.} 106, 21599. doi:10.1029/2000JA000238

\bibitem[McCracken et al.(2013)]{McCracken2013} McCracken, K. G.,
Beer, J., Steinhilber, F., and Abreu, J. (2013). A phenomenological
study of the cosmic ray variations over the past 9400 years, and their
implications regarding solar activity and the solar dynamo. \emph{Solar
Physics} 286, 609. doi:10.1007/s11207-013-0265-0

\bibitem[Morice et al.(2012)]{Morice}Morice, C. P., Kennedy, J. J.,
Rayner, N. A., and Jones, P. D. (2012). Quantifying uncertainties
in global and regional temperature change using an ensemble of observational
estimates: the HadCRUT4 dataset. \emph{J. Geophys. Res.} 117, D08101.

\bibitem[Mörner et al.(2015)]{Morner(2015)}Mörner, N.-A., Scafetta,
N., Solheim, J.-E. (2015). The January 7 Giant Solar Flare, the Simultaneous
Triple Planetary Conjunction and Additional Records at Tromsø, Northern
Norway. In ``\emph{Planetary Influence on the Sun and the Earth,
and a Modern Book-Burning}.'' pp. 33--38, Nova Science Publisher,
New York. ISBN: \href{https://novapublishers.com/shop/planetary-influence-on-the-sun-and-the-earth-and-a-modern-book-burning/}{9781634828376}

\bibitem[Mitalas and Sills(1992)]{Mitalas} Mitalas, R., and Sills,
K. (1992). On the photon diffusion time scale for the sun. The Astrophysical
Journal 401, 759--760.

\bibitem[Moons and Morbidelli(1995)]{Moons}Moons, M., and Morbidelli,
A. (1995). Secular resonances inside mean-motion commensurabilities:
the 4/1, 3/1, 5/2 and 7/3 cases. \emph{Icarus} 114, 33--50.

\bibitem[NASA(2014)]{NASA2014a} NASA News (2014). Sun unleashes first
X-class flare of 2014. Jan. 7. Available at: \href{http://svs.gsfc.nasa.gov/vis/a010000/a011100/a011136/}{http://svs.gsfc.nasa.gov/vis/a010000/a011100/a011136/}

\bibitem[Neff et al.(2001)]{Neff}Neff, U., Burns, S. J., Mangini,
A., Mudelsee, M., Fleitmann, D. and Matter, A. (2001). Strong coherence
between solar variability and the monsoon in Oman between 9 and 6
kyear ago. \textit{Nature} 411, 290.doi:10.1038/35077048

\bibitem[Ogurtsov et al.(2002)]{Ogurtsov} Ogurtsov, M. G., Nagovitsyn,
Y. A., Kocharov, G. E., and Jungner, H. (2002). Long-period cycles
of the sun's activity recorded in direct solar data and proxies. \emph{Solar
Physics} 211, 371--394.

\bibitem[Parker(1955)]{Parker1955}Parker, E. N. (1955). Hydromagnetic
Dynamo Models. \emph{Astrophysical Journal}, 122, 293--314.

\bibitem[Parker(1958)]{Parker}Parker, E. N. (1958). Dynamics of the
Interplanetary Gas and Magnetic Fields. \emph{Astrophysical Journal},
128, 664-676.

\bibitem[Petrakou(2021)]{Petrakou}Petrakou, E. (2021). Planetary
statistics and forecasting for solar flares. \emph{Advances in Space
Research} 68, 2963--2973. doi: 10.1016/j.asr.2021.05.034

\bibitem[Pikovsky et al.(2001)]{Pikovsky} Pikovsky, A., Rosemblum,
M., Kurths, J. (2001). \emph{Synchronization: A Universal Concept
in Nonlinear Sciences}. Cambridge University Press.

\bibitem[Press et al.(1997)]{Press}Press, W. H., Teukolsky, S. A.,
Vetterling, W. T., and Flannery, B. P. (1997). \emph{Numerical Recipes
in C, 2nd Edn.}, Cambridge University Press.

\bibitem[Reimer et al.(2004)]{Reimer}Reimer, P. J., Baillie, M. G.
L., Bard, E., \textit{\emph{et al.}} (2004). Intcal04 terrestrial
radiocarbon age calibration, 0-26 cal kyear BP. \textit{Radiocarbon}
46, 1029. doi:10.1017/S0033822200032999. Available at \href{https://www.radiocarbon.org/IntCal04.htm}{https://www.radiocarbon.org/IntCal04.htm}

\bibitem[Salvador(2013)]{Salvador}Salvador, R. J. (2013). A mathematical
model of the sunspot cycle for the past 1000 year. \emph{Pattern Recogn.
Phys.} 1, 117--122. doi:10.5194/prp-1-117-2013.

\bibitem[Scafetta et al.(2004)]{Scafetta2004}Scafetta, N., Grigolini,
P., Imholt, T., Roberts, J. A., and West, B. J. (2004). Solar turbulence
in earth\textquoteright s global and regional temperature anomalies.
Physical Review E, 69, 026303.

\bibitem[Scafetta et al.(2006)]{Scafetta2006}Scafetta, N., and West,
B. J. (2006). Phenomenological solar signature in 400 years of reconstructed
Northern Hemisphere temperature record. Geophysical Research Letters,
33, L17718.

\bibitem[Scafetta(2009)]{Scafetta2009}Scafetta, N. (2009). Empirical
analysis of the solar contribution to global mean air surface temperature
change. \textit{Journal of Atmospheric and Solar-Terrestrial Physics}
71, 1916--1923.

\bibitem[Scafetta(2010)]{Scafetta2010} Scafetta, N. (2010). Empirical
evidence for a celestial origin of the climate oscillations and its
implications. \textit{Journal of Atmospheric and Solar-Terrestrial
Physics} 72, 951--970. doi:10.1016/j.jastp.2010.04.015, 2010.

\bibitem[Scafetta et al.(2011)]{Scafetta2011}Scafetta, N. (2011).
Total solar irradiance satellite composites and their phenomenological
effect on climate. In \emph{Evidence-Based Climate Science}. (Easterbrook
D., Elsevier), 12, 289--316.

\bibitem[Scafetta(2012a)]{Scafetta2012a} Scafetta, N. (2012a). Multi-scale
harmonic model for solar and climate cyclical variation throughout
the Holocene based on Jupiter-Saturn tidal frequencies plus the 11
year solar dynamo cycle. \textit{Journal of Atmospheric and Solar-Terrestrial
Physics} 80, 296--311. doi:10.1016/j.jastp.2012.02.016, 2012a.

\bibitem[Scafetta(2012b)]{Scafetta2012b}Scafetta, N. (2012b). Does
the Sun work as a nuclear fusion amplifier of planetary tidal forcing?
A proposal for a physical mechanism based on the mass-luminosity relation.
\textit{Journal of Atmospheric and Solar-Terrestrial Physics} 81--82,
27--40. doi:10.1016/j.jastp.2012.04.002.

\bibitem[Scafetta(2012c)]{Scafetta2012c}Scafetta, N. (2012c). A shared
frequency set between the historical mid-latitude aurora records and
the global surface temperature. \textit{Journal of Atmospheric and
Solar-Terrestrial Physics} 74, 145--163.

\bibitem[Scafetta(2012d)]{Scafetta2012d}Scafetta, N. (2012d). Testing
an astronomically based decadal-scale empirical harmonic climate model
versus the IPCC (2007) general circulation climate models. \textit{Journal
of Atmospheric and Solar-Terrestrial Physics} 80, 124--137.

\bibitem[Scafetta(2013)]{Scafetta2013}Scafetta, N. (2013). Discussion
on climate oscillations: CMIP5 general circulation models versus a
semi-empirical harmonic model based on astronomical cycles. \emph{Earth-Sci.
Rev.} 126, 321--357.

\bibitem[Scafetta and Willson(2013a)]{ScafettaWillson2013a} Scafetta,
N., and Willson, R. C. (2013a). Planetary harmonics in the historical
Hungarian aurora record (1523--1960). \emph{Planet. Space Sci.} 78,
38--44. doi:10.1016/j.pss.2013.01.005, 2013a.

\bibitem[Scafetta and Willson(2013b)]{ScafettaWillson2013b} Scafetta,
N., and Willson, R. C. (2013b). Empirical evidences for a planetary
modulation of total solar irradiance and the TSI signature of the
1.09 year Earth-Jupiter conjunction cycle. \emph{Astrophys. Space
Sci.} 348, 25--39. doi:10.1007/s10509-013-1558-3.

\bibitem[Scafetta and Willson(2013c)]{ScafettaWillson2013c} Scafetta,
N., and Willson, R. C. (2013c). Multiscale comparative spectral analysis
of satellite total solar irradiance measurements from 2003 to 2013
reveals a planetary modulation of solar activity and its nonlinear
dependence on the 11 year solar cycle. \emph{Pattern Recognition in
Physics} 1, 123--133. doi:10.5194/prp-1-123-2013.

\bibitem[Scafetta(2014a)]{Scafetta2014a}Scafetta, N. (2014a). The
Complex Planetary Synchronization Structure of the Solar System. \emph{Pattern
Recognition in Physics} 2, 1--19. doi:10.5194/prp-2-1-2014

\bibitem[Scafetta(2014b)]{Scafetta2014b}Scafetta, N. (2014b). Discussion
on the spectral coherence between planetary, solar and climate oscillations:
a reply to some critiques. \textit{Astrophysics and Space Science},
354, 275--299. doi:10.1007/s10509-014-2111-8

\bibitem[Scafetta(2014c)]{Scafetta2014c}Scafetta, N. (2014c). Multi-scale
dynamical analysis (MSDA) of sea level records versus PDO, AMO, and
NAO indexes. \emph{Climate Dynamics} 43, 175--192.

\bibitem[Scafetta(2016)]{Scafetta2016}Scafetta, N. (2016). High resolution
coherence analysis between planetary and climate oscillations. \textit{Advances
in Space Research}\textit{\emph{ 57(10), 2121--2135.}}

\bibitem[Scafetta et al.(2016)]{Scafettaetal2016} Scafetta, N., Milani,
F., Bianchini, A., and Ortolani, S. (2016). On the astronomical origin
of the Hallstatt oscillation found in radiocarbon and climate records
throughout the Holocene. \textit{Earth-Sci. Rev.} 162, 24. doi:10.1016/j.earscirev.2016.09.004

\bibitem[Scafetta(2018)]{Scafetta2018}Scafetta, N. (2018). Reply
on Comment on \textquotedblleft High resolution coherence analysis
between planetary and climate oscillations\textquotedblright{} by
S. Holm.\textit{ Advances in Space Research}\textit{\emph{ 62, 334--342.}}

\bibitem[Scafetta et al.(2019a)]{Scafetta2019}Scafetta, N., Milani,
F., Bianchini, A. (2019a). Multiscale Analysis of the Instantaneous
Eccentricity Oscillations of the Planets of the Solar System from
13,000 BC to 17,000 AD.\textit{ }\emph{Astronomy Letters} 45(11),
778--790.

\bibitem[Scafetta et al.(2019b)]{Scafetta2019b}Scafetta, N., Willson,
R .C., Lee, J. N., and Wu, D. L. (2019b). Modeling quiet solar luminosity
variability from TSI satellite measurements and proxy models during
1980-2018.\textit{ }\emph{Remote Sensing}, 11(21), 2569.

\bibitem[Scafetta et al.(2020)]{Scafettaetal2020}Scafetta, N., Milani,
F., Bianchini, A. (2020). A 60-Year Cycle in the Meteorite Fall Frequency
Suggests a Possible Interplanetary Dust Forcing of the Earth's Climate
Driven by Planetary Oscillations.\textit{ }\emph{Geophysical Research
Letters}, 47(18), e2020GL089954.

\bibitem[Scafetta(2020)]{Scafetta2020}Scafetta, N. (2020). Solar
Oscillations and the Orbital Invariant Inequalities of the Solar System.
\emph{Solar Physics} 295, 33. doi:10.1007/s11207-020-01599-y

\bibitem[Scafetta(2021)]{Scafetta2021}Scafetta, N. (2021). Reconstruction
of the Interannual to Millennial Scale Patterns of the Global Surface
Temperature.\textit{ }\emph{Atmosphere}, 12, 147.

\bibitem[Schwabe(1843)]{Schwabe}Schwabe H. (1843). Sonnenbeobachtungen
im Jahre 1843\textquotedbl{} {[}Observations of the sun in the year
1843{]}. Astronomische Nachrichten 21, 233--236.

\bibitem[Shkolnik et al.(2003)]{Shkolnik2003}Shkolnik, E., Walker,
G. A. H., and Bohlender D. A. (2003). Evidence for Planet-induced
Chromospheric Activity on HD 179949. \emph{The Astrophysical Journal}
556, 296--301.

\bibitem[Shkolnik et al.(2005)]{Shkolnik2005}Shkolnik, E., Walker,
G. A. H., Bohlender, D. A., Gu P.-G., and Kurster, M. (2005). Hot
Jupiters and Hot Spots: The Short- and Long-Term Chromospheric Activity
on Stars with Giant Planets. \emph{The Astrophysical Journal} 622,
1075--1090.

\bibitem[Smythe and Eddy(1977)]{Smythe}Smythe, C. M., Eddy, J. A.
(1977). Planetary tides during Maunder sunspot. \emph{Nature} 266,
434--435.

\bibitem[Solheim(2013)]{Solheim}Solheim, J.-E. (2013). The sunspot
cycle length -- modulated by planets? \emph{Pattern Recognition in
Physics} 1, 159--164. doi:10.5194/prp-1-159-2013.

\bibitem[Stefani et al.(2016)]{Stefani2016}Stefani, F., Weber, N.,
Weier, T., and Giesecke, A. (2016). Synchronized helicity oscillations:
A link between planetary tides and the solar cycle? \emph{Solar Physics}
291, 2197.

\bibitem[Stefani(2018)]{Stefani2018}Stefani, F., Giesecke, A., Weber,
N., Weier, T. (2018). On the synchronizability of Tayler-Spruit and
Babcock-Leighton type dynamos. \emph{Solar Physics} 293, 12.

\bibitem[Stefani et al.(2019)]{Stefani2019}Stefani, F., Giesecke,
A., Weier, T. (2019). A model of a tidally synchronized solar dynamo.
\emph{Solar Physics} 294, 60.

\bibitem[Stefani et al.(2020a)]{Stefani(2020)}Stefani, F., Beer,
J., Giesecke, A., et al. (2020a). Phase coherence and phase jumps
in the Schwabe cycle. \emph{Astron. Nachr.} 20, 341, 600--615.

\bibitem[Stefani et al.(2020b)]{Stefani2020b}Stefani, F., Giesecke,
A., Seilmayer, M., Stepanov, R., Weier T. (2020b). Schwabe, Gleissberg,
Suess-de Vries: Towards a consistent model of planetary synchronization
of solar cycles. \emph{Magnetohydrodynamics} 56, No. 2/3, 269-280.

\bibitem[Stefani et al.(2021)]{Stefani2021}Stefani, F., Stepanov,
R., and Weier, T. (2021). Shaken and Stirred: When Bond Meets Suess-De
Vries and Gnevyshev-Ohl. \emph{Sol. Phys.} 296, 88.

\bibitem[Stephenson(1974)]{Stephenson1974}Stephenson, B. (1974).\emph{
The Music of the Heavens.} Princeton University Press, Princeton,
NJ.

\bibitem[Steinhilber et al.(2009)]{Steinhilber} Steinhilber F., Beer
J., and Fröhlich C. (2009). Total solar irradiance during the Holocene.
\emph{Geophysics Research Letters} 36, L19704.

\bibitem[Steinhilber et al.(2012)]{Steinhilber2012}Steinhilber, F.,
Abreu, J. A., Beer, J., Brunner, I., Christl, M., Fischer, H., Heikkila,
U., Kubik, P. W., Mann, M., McCracken, K. G., Miller, H., Miyahara,
H., Oerter, H., Wilhelms, F. (2012). 9,400 years of cosmic radiation
and solar activity from ice cores and tree rings. \emph{PNAS} 109(16),
5967--5971.

\bibitem[Stix(2003)]{Stix} Stix, M. (2003). On the time scale of
energy transport in the sun. \emph{Solar Physics} 212, 3--6.

\bibitem[Stuiver et al.(1998)]{Stuiver1998}Stuiver, M., Reimer, P.
J., Bard, E., Beck, J. W., Burr, G. S., Hughen, K. A., Kromer, B.,
McCormac, G., Van Der Plicht, J. and Spurk, M. (1998). INTCAL98 Radiocarbon
age calibration, 24,000-0 cal BP. \textit{Radiocarbon} 40, 1041. doi:10.1017/S0033822200019123

\bibitem[Suess(1965)]{Suess1965}Suess H. E. (1965). Secular variations
of the cosmic-ray-produced carbon 14 in the atmosphere and their interpretations.
\emph{J. Geophys. Res.} 70, 5937--5952.

\bibitem[Tattersall(2013)]{Tattersall}Tattersall, R. (2013). The
Hum: Lognormal distribution of Planetary-Solar resonance. \emph{Pattern
Recognition in Physics} 1, 185--198.

\bibitem[Tobias(2002)]{Tobias2002}Tobias, S. M. (2002). The solar
dynamo. \emph{Philosophical Transactions on the Royal Society A} 360(1801),
2741--2756.

\bibitem[Vos et al.(2004)]{Vos}Vos, H., Brüchmann, C., Lücke, A.,
Negendank, J.F.W., Schleser, G.H., Zolitschka, B. (2004). Phase stability
of the solar Schwabe cycle in lake Holzmaar, Germany, and GISP2, Greenland,
between 10,000 and 9,000 cal. BP. In: Fischer, H., Kumke, T., Lohmann,
G., Flöser, G., Miller, H., von Storch, H., Negendank, J.F. (eds.),
\emph{The Climate in Historical Times: Towards a Synthesis of Holocene
Proxy Data and Climate Models}. GKSS School of Environmental Research,
Springer, Berlin, 293.

\bibitem[Wagner et al.(2001)]{Wagner2001} Wagner, G., Beer, J., Masarik,
J., Muscheler, R., Kubik, P. W., Mende, W., Laj, C., Raisbeck, G.
M., Yiou, F. (2001). Presence of the solar de Vries cycle (\ensuremath{\sim}205
years) during the last ice age. \emph{Geophys. Res. Lett.} 28, 303-306.

\bibitem[Weiss and Tobias(2016)]{Weiss(2016)}Weiss, N.O., Tobias,
S.M. (2016). Supermodulation of the Sun\textquoteright s magnetic
activity: the effects of symmetry changes. \emph{Mon. Not. Roy. Astron.
Soc.} 456, 2654--2661.

\bibitem[Wilson(2013)]{Wilson}Wilson, I. R. G. (2013). The Venus--Earth--Jupiter
spin--orbit coupling model, \emph{Pattern Recognition in Physics}
1, 147--158, doi:10.5194/prp-1-147-2013.

\bibitem[Willson and Mordvinov(1999)]{Willson1999} Willson, R. C.,
amd Mordvinov, V. (1999). Time-Frequency Analysis of Total Solar Irradiance
Variations. \emph{Geophys. Res. Lett.} 26(24), 3613--3616. doi:10.1029/1999GL010700.

\bibitem[Willson and Mordvinov(2003)]{Willson2003} Willson, R. C.,
and Mordvinov, V. (2003). Secular total solar irradiance trend during
solar cycles 21-23. \emph{Geophys. Res. Lett.} 30, 1199. doi:10.1029/2002GL016038.

\bibitem[Wolf(1859)]{Wolf} Wolf, R. (1859). Extract of a letter to
Mr. Carrington. \emph{Monthly Notices of the Royal Astronomical Society}
19, 85--86.

\bibitem[Wyatt and Curry(2014)]{Wyatt}Wyatt, M. G., and Curry, J.
A. (2014). Role for Eurasian Arctic shelf sea ice in a secularly varying
hemispheric climate signal during the 20th century. \emph{Climate
Dynamics} 42, 2763--2782.

\bibitem[Wood(1986)]{Wood}Wood, F. J. (1986). \emph{Tidal Dynamics,
}Reidel, Dordrecht, the Netherlands.

\bibitem[Wolff and Patrone(2010)]{Wolff} Wolff, C. L., and Patrone,
P. N. (2010). A new way that planets can affect the Sun. \emph{Solar
Physics} 266, 227. doi:10.1007/s11207-010-9628-y

\bibitem[Yamaguchia et al.(2010)]{Yamaguchia} Yamaguchia Y., Yokoyamaa
Y., Miyaharad H., Shoe K., and Nakatsukaf T., (2010). Synchronized
Northern Hemisphere climate change and solar magnetic cycles during
the Maunder Minimum. \emph{PNAS} 107, 20697--20702.

\bibitem[Yu et al.(1983)]{Yu1983} Yu, Z., Chang, S., Kumazawa, M.,
Furumoto, M., and Yamamoto, A. (1983). Presence of periodicity in
meteorite falls. \emph{National Institute of Polar Research, Memoirs},
Special Issue 30, 362--366.

\bibitem[Zaqarashvili et al.(2010)]{Zaqarashvili2010}Zaqarashvili,
T. V., Carbonell, M., Oliver, Ballster, J. L. (2010). Magnetic Rossby
waves in the solar tachocline and Rieger-type periodicities. \emph{The
Astrophysical Journal} 709, 749--758.

\bibitem[Zaqarashvili(2018)]{Zaqarashvili(2018)}Zaqarashvili, T.
V. (2018). Equatorial magnetohydrodynamic shallow water waves in the
solar tachocline. \emph{Astrophys. J.} 856, 32.

\bibitem[Zaqarashvili et al.(2021)]{Zaqarashvili}Zaqarashvili, T.
V., Albekioni, M., Ballester, J. L. et al. (2021). Rossby Waves in
Astrophysics. \emph{Space Sci Rev} 217, 15.

\bibitem[Zioutas et al.(2022)]{Zioutas}Zioutas, K., Maroudas, M.,
Kosovichev, A. (2022). On the Origin of the Rhythmic Sun\textquoteright s
Radius Variation. \emph{Symmetry} 14, 325.
\end{thebibliography}
\end{document}